\documentclass{article}

\usepackage{amsfonts} 
\usepackage{amssymb,amsmath} 
\usepackage{graphicx,color}
 \usepackage[breaklinks,colorlinks=true,linkcolor=blue,citecolor=red,backref=page]{hyperref}

\DeclareMathAlphabet{\itbf}{OML}{cmm}{b}{it}

\def\bq{{{\itbf q}}}

\def\br{{{\itbf r}}}
\def\bs{{{\itbf s}}}
\def\by{{{\itbf y}}}
\def\bx{{{\itbf x}}}

\def\bk{{{\itbf k}}}

 \def\bzeta{{\boldsymbol{\zeta}}}
\def\bxi{{\boldsymbol{\xi}}}

\def\eps{{\varepsilon}}

\newcommand{\RR}{\mathbb{R}}

\newcommand{\EE}{\mathbb{E}}

 \newtheorem{thm}{Theorem}[section]

\newtheorem{prop}[thm]{Proposition}

\newcommand{\ea}{\end{eqnarray}}  
\newcommand{\ba}{\begin{eqnarray}}  
\newcommand{\ee}{\end{equation}}  
\newcommand{\be}{\begin{equation}}  
\newcommand{\ean}{\end{eqnarray*}}  
\newcommand{\ban}{\begin{eqnarray*}}  
\begin{document}

 \title{Scintillation of Partially Coherent Light in Time Varying Complex Media}

\author{Josselin
Garnier\thanks{\footnotesize Centre de Math\'ematiques Appliqu\'ees, Ecole Polytechnique,
91128 Palaiseau Cedex, France (josselin.garnier@polytechnique.edu)} 
\and Knut S\o lna\thanks{\footnotesize Department of Mathematics, 
University of California, Irvine CA 92697
(ksolna@math.uci.edu)}
}

% \author{
%\name{Josselin Garnier\textsuperscript{a}  and Knut S\o lna\textsuperscript{b}}
%\affil{\textsuperscript{a}Centre de Math\'ematiques Appliqu\'ees, Ecole Polytechnique, 91128 Palaiseau Cedex,
%France
%{\tt josselin.garnier@polytechnique.edu}; \textsuperscript{b}Department of Mathematics, 
%University of California, Irvine CA 92697
%{\tt ksolna@math.uci.edu}}
%}
% 

\maketitle

%\author{Josselin Garnier\authormark{1,3} and Knut S\o lna\authormark{2,4}}
%
%\address{\authormark{1}Centre de Math\'ematiques Appliqu\'ees, Ecole Polytechnique, Institut Polytechnique de Paris, 91128 Palaiseau Cedex,
%France  \\
%\authormark{2}Department of Mathematics, 
%University of California, Irvine CA 92697, USA
%  \\
%\authormark{3} josselin.garnier@polytechnique.edu \\
%\authormark{4} ksolna@math.uci.edu
%}

%\email{\authormark{*}opex@osa.org} %% email address is required



\begin{abstract}
We present a theory for  wave scintillation in the situation with a time-dependent partially coherent source and a time-dependent randomly heterogeneous medium.  
Our objective is to understand how the scintillation index of the measured intensity depends on the source and medium parameters.
%multiscale wave motion setting. 
%The wave field in the time-dependent medium is  characterized via an It\^o-type diffusion equation.
We deduce from an asymptotic analysis of the random wave equation a general form of the scintillation index 
and we evaluate this  in various scaling regimes.
The scintillation index is a fundamental quantity that is used  to analyze and optimize imaging and communication schemes. 
 Our results are useful to quantify the scintillation index under realistic propagation scenarios and to address such 
  optimization challenges. 
\end{abstract}
 
 \newpage
 
 \tableofcontents  
 
%\setboolean{copyright}{true} 
%%%%%%%%%%%%%%%%%%%%%%%%%%  body  %%%%%%%%%%%%%%%%%%%%%%%%%%
\section{Introduction}
  
 We consider the fundamental problem of characterizing the scintillation of optical measurements
with a time-dependent partially coherent source and a time-dependent random medium.  
   The scintillation index corresponds to a measure 
 of the signal-to-noise ratio or relative strength of fluctuations in the intensity.
 If $I$ is the measured intensity (irradiance)  then we define the scintillation index by 
 \begin{equation}
{\cal S} = \frac{ \EE[I^2] - \EE[I]^2
 }
{
 \EE[I]^2
}  ,
\end{equation}
where $\EE[\cdot]$ stands for the statistical expectation obtained by averaging over repeated measurements 
under  independent and identically distributed conditions.
Modeling and analysis of laser speckle and scintillation is a classic challenge in optics \cite{dainty,tatarskii,andrews}.
 A rigorous mathematical analysis and quantification of scintillation  has been a long standing open question
 despite the long history and importance of this challenge.  General insight about  what governs the scintillation 
is important in the design of optical systems, for instance  for imaging and communication
 through the  turbulent atmosphere \cite{baykal} and through  oceanic turbulence \cite{xu}.
 The challenge of choosing appropriate sources for scintillation control 
 has received a lot of attention \cite{gbur}. 

In \cite{gar14a,gar16a} we presented
 an analysis of the scintillation problem for deterministic coherent beams and plane wave sources and time-independent media.
 In this paper 
we consider the scintillation problem when the source is partially coherent in time and space
and the medium has time and space random fluctuations.
Partially coherent sources 
have indeed been promoted for reducing scintillation at a receiving end 
in the context of laser propagation \cite{svetlana,svetlana2,xiao,gbur}.
Most of these studies rely on physical experiments or
  numerics and Monte Carlo simulations to evaluate the scintillation index. 
  From the theoretical point of view, 
  analysis of wave propagation can be  carried out in a perturbative regime using in particular 
  Rytov theory with small fluctuations in the wave field to obtain insight
  about the scintillation \cite{shan,baker}. 
 The fluctuations of intensity over different receiver response times have been studied in \cite{fante79,banach1,banach2} in the limits of very slow or very fast detectors. 
 The effect of temporal coherence on scintillation for weak turbulence was considered in \cite{fante81}.
   An analysis of  scintillation  and how it depends on the smoothness of deterministic initial condition 
  is  presented in \cite{bal} with a focus on understanding 
  self-averaging situations with a vanishing scintillation index.  Issues related to aperture averaging 
 is also 
 considered  in  \cite{shaw} in the context of deterministic sources by using a path integral approach 
 for modeling the effect of turbulence.

Here we consider
the high-frequency and far-field situation where the effect of the random medium can be captured by a white-noise term in the  It\^o-Schr\"odinger equation that governs the evolution of the wave field \cite{char1,char2}.
This equation can address both weak intensity fluctuations and strong intensity fluctuations (the saturated regime).
The response time of the photodetector, the coherence times of the source and the random medium can be arbitrary, provided that they are larger than the travel time of the field from the source to the detector through the medium.
%We consider the saturated regime where 
%the incoherent fluctuations of a wave field generated by a coherent souce can be of the same order as or larger than the coherent field.
Under such circumstances,
the effective reduced system (\ref{eq:fouriermom0eta})  for the fourth-order moments of the wave field can be derived from the  It\^o-Schr\"odinger equation and used
for numerical evaluation of the scintillation index in the general high-frequency  and far-field situation. 
Based on this system
we  obtain explicit expressions of the scintillation index in three scaling regimes
determined by the ratio of the correlation radius of the source over the correlation radius of the medium.
We   then explicitly  characterize
scintillation with partially coherent sources and time-dependent random media
and quantify how the space-time statistical parameters of the source and medium  affect the scintillation index. 
 An important aspect of our analysis is that we allow the medium to be time-dependent, so that it changes on a time scale that is slow relative to the travel time of the optical field.  This is the situation  in the context of 
 laser beam propagation in the atmosphere with turbulence creating slow temporal changes of the medium.  The detector in our modeling has  a finite 
 response time  which can be on the time scale of the changes in the medium.
 The averaging at the detector can have a strong impact on the scintillation index depending
 on the characteristic  time scales involved.

\begin{figure}[h!]
\begin{center}
\begin{picture}(330,80)
 \put(10,0){\includegraphics[width=7.7cm]{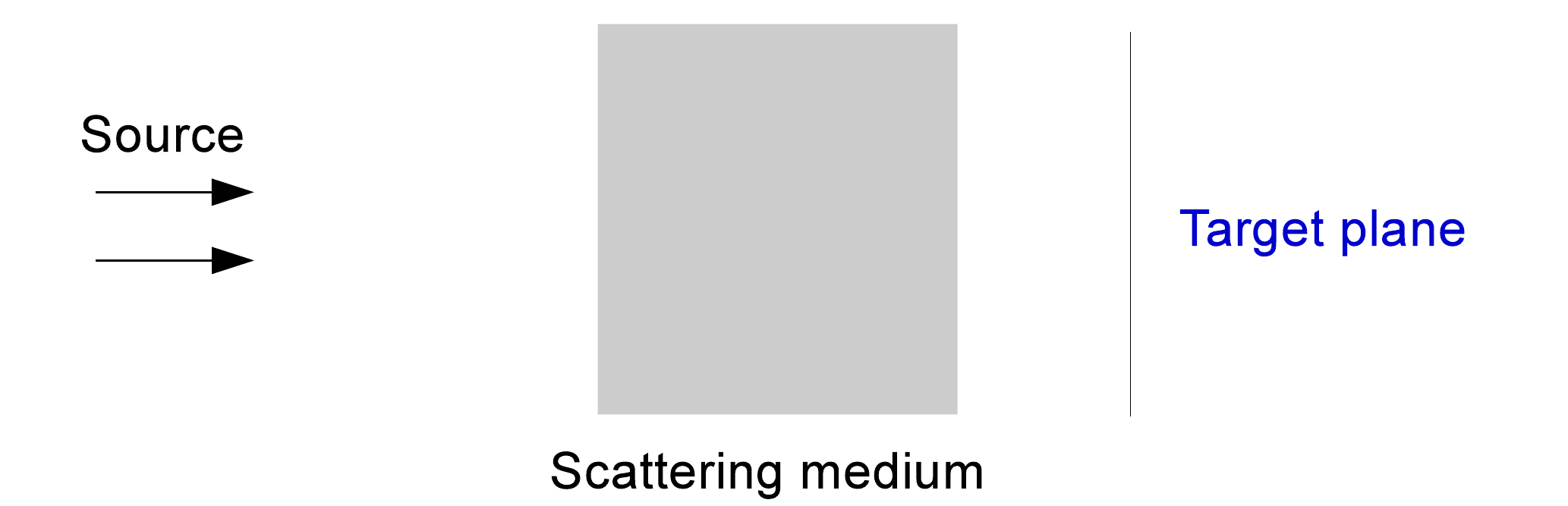}}
%\put(155,4){\includegraphics[width=3.3cm]{figure_focusing_speck1.eps}}
%\put(195,4){\includegraphics[width=2.3cm]{fig_scint2.jpg}}
\put(225,4){\includegraphics[width=2.3cm]{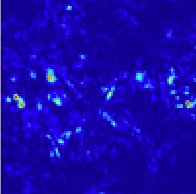}}
 \end{picture}
\end{center}
\caption{The figure shows the configuration that we   consider. 
A partially coherent source fluctuating randomly in space and time is impinging on a complex medium.  
The complex medium is modeled as random and changes
in time. The time changes in the medium happen on the recording time scale of the detector,
but are slow relative to the travel time of the wave over the considered range. 
Due to time averaging 
the detector measures a smoothed version of the intensity and we seek to characterize 
the  scintillation index of this measurement, which corresponds to a  signal-to-noise ratio. 
}\label{fig:1}
 \end{figure}

     The configuration that we consider is illustrated in Figure~\ref{fig:1} with a 
   partially coherent source field impinging from the left and propagating through a 
   random  medium and then the scintillating intensity pattern is recorded at the receiver end. 
   Our objective is to characterize the scintillation index of the observed transmitted intensity pattern
   shown to the right in the figure.    
  
 We comment on a special, but important,  case  corresponding to the wave  field having a Gaussian distribution. 
 Indeed it is a well-accepted conjecture that the statistics of the complex  wave field becomes 
circularly  symmetric complex  
Gaussian when the wave propagates through the turbulent atmosphere \cite{valley,ya}
and the conjecture  can be proved  in certain situations \cite{book1,lenya,gu21}. 
 In the Gaussian case   the intensity   is the sum of the squares of two independent Gaussian random variables, which 
up to a scaling  has 
$\chi$-square distribution with two degrees of freedom, that is, 
an exponential distribution.  This situation gives a  unit scintillation index. 
Based on an analysis of the fourth moment  of the wave field we identify in this paper regimes 
 that correspond to a unit value for the scintillation index and  which 
are consistent with the Gaussian conjecture. 
The case with a Gaussian field distribution  is the critical situation with the  signal-to-noise ratio  of  the intensity being one. 
In general the scintillation index can be  below one for small fluctuations 
 in the intensity and can reach values beyond one when the intensity distribution has heavier tails 
 than those corresponding to the exponential distribution. 
 We encounter both situations in this paper and discuss what type of scaling regime
 may lead to such situations.

The outline of the paper is as follows. We formulate the problem in Section
\ref{sec:1}. This involves defining the statistical models
for the source and the medium and deriving the stochastic partial differential
equation, the It\^o-Schr\"odinger equation,    that  characterizes the wave field.
We then  relate the solution of the stochastic partial differential
equation to
the  measured scintillation index.  
The main theoretical foundation  for  analyzing the scintillation is  a  framework  for analyzing the  fourth-order moment of the wave field 
and we discuss this in Section \ref{sec:4}. 
In Section \ref{sec:main} we give the main results
which characterize the scintillation index  in various scaling regimes.  
In Section \ref{sec:S} we present an example involving data presented 
in \cite{svetlana}.  Technical calculations associated with the fourth-order  
moment equations are presented in the appendices. 

\section{Probing Time-Dependent Complex Media with Partially Coherent Sources}
\label{sec:1}%

In this section we outline the modeling and the problem that we will consider. 
In section \ref{sec:i1} we describe the statistical modeling of the source and of the
random medium.
In section \ref{sec:i2} we give the It\^o-Schr\"odinger equation 
that describes the evolution of the wave field in the random medium.
In section \ref{sec:i3} we relate the random transmitted wave  field    
to the quantity of interest which is the scintillation index of the measurements.  

\subsection{Source and Medium Modeling}
\label{sec:i1}%
The time-harmonic field $U(z,\bx,t)$ satisfies the Helmholtz equation
\begin{equation}
\label{eq:helm}
 \Delta {U} +
k_o^2 {\rm n}^2(z,\bx,t)   {U}  = - \delta(z) f(\bx,t) ,\quad \quad (z,\bx) \in \RR \times \RR^2,
\end{equation}
%with radiation conditions and 
where
$k_o=2\pi/\lambda_o$ is the central wavenumber ($\lambda_o$ is the central wavelength). 
Here $t$ is the slow time corresponding to the 
time  at which the random medium and the source change.
The coherence times of the medium and source are assumed to be much larger than the travel time from the source to the detector through the medium so that $t$  is a frozen parameter in  (\ref{eq:helm}).

The source $f$ in the plane $z=0$ is partially coherent,
statistically stationary in space and time.
We model it as a complex Gaussian process
with mean zero, variance one, and covariance 
\begin{equation}
\EE \Big[  f\big(  \bx+\frac{\by}{2} , t+\frac{\tau}{2} \big)
 \overline{f} \big(   \bx-\frac{\by}{2},  t-\frac{\tau}{2}\big)  \Big] =
 F\left(\frac{\tau}{\tau_{\rm s}}\right) 
 \exp \Big( -\frac{|\by|^2}{4 \ell_{\rm s}^2}  \Big)  , \label{eq:c1}
\end{equation}
where $\ell_{\rm s}$, resp. $\tau_{\rm s}$,  is the correlation radius, resp. the coherence time, of the source 
 and the time covariance function $F$ is normalized so that $F(0)=1$ and $\int_0^\infty F(s) ds=1$. 
 Here the correlation radius of the source $\ell_{\rm s}$ is assumed to be small relative to the range $L$
 and
the coherence time  $\tau_{\rm s}$ is assumed to be much larger than the propagation time $L/c_o$,
where $c_o$ is the background speed of propagation and $L$ is the distance from the source to the detector.
%That is why the time $t$ is frozen in the Helmholtz equation (\ref{eq:helm}).
For convenience we use here a Gaussian correlation function for the spatial source correlations,
but we remark that we could have used a more general form.  
A more detailed model for the source, in particular a discussion of realization via 
Spatial Light Modulators (SLMs)  can be found in \cite{garniers5,svetlana2}.  
For other approaches to the generation of the partially coherent source we refer to \cite{voelz}  for instance.

The medium is  random  and we denote by  $\nu$ the 
relative fluctuations   
in the square index of refraction: ${\rm n}^2 (z,\bx,t)=1+\nu(z,\bx,t)$.
The stochastic process $\nu$ 
is stationary in space and time and zero-mean, and its covariance function is of the form
\begin{equation}
\label{eq:nu}
\EE [ \nu(z'+  z,\bx'+  \bx,t+\tau) \nu(z',\bx',t) ] = 
\sigma_{\rm m}^2 G\left( \frac{\tau}{\tau_{\rm m}}\right) 
{\cal C}_{\rm m}\left( \frac{z}{\ell_{\rm m}},\frac{\bx}{\ell_{\rm m}}\right)  ,
\end{equation}
where $\ell_{\rm m}$, resp. $\tau_{\rm m}$, is the correlation radius, resp. the coherence time, of the random medium fluctuations, $\sigma_{\rm m}$ is the standard deviation of the fluctuations of the square index of refraction, and the functions $G$ and ${\cal C}_{\rm m}$ are normalized so that $G(0)=1$, $\int_0^\infty G(s) ds=1$, ${\cal C}_{\rm m}(0,{\bf 0})=1$, $\int_\RR
{\cal C}_{\rm m}(\zeta,{\bf 0}) d\zeta=1$,  and $\int_{\RR^2}
{\cal C}_{\rm m}(0,\boldsymbol{\chi}) d\boldsymbol{\chi}=1$.
%We denote $C(\boldsymbol{\chi}) = \int_\RR {\cal C}_{\rm m}(\zeta,\boldsymbol{\chi}) d\zeta$, which is such that $C({\bf 0})=1$
% so that we can write
%\be 
% \int_\RR \EE[\nu(z'+  z,\bx'+  \bx,t+\tau) \nu(z',\bx',t)]   dz 
% =
%\sigma_{\rm m}^2 \ell_{\rm m} G\left( \frac{\tau}{\tau_{\rm m} } \right)  C\left( \frac{\bx}{\ell_{\rm m} } \right) 
%  . \label{eq:c2} 
%\ee
The special case where the correlation function corresponds to Kolmogorov turbulence is discussed in \cite{garF}. 
Here the correlation radius of the random medium $\ell_{\rm m}$ is assumed to be small relative to the range $L$
and 
the coherence time  of the medium $\tau_{\rm m}$  is
assumed to be much larger than 
the propagation time $L/c_o$. Thus,  the `turnover time'  of the medium is long compared 
to the propagation time,  however,  we assume that it is on the scale of the coherence time of the source $\tau_{\rm s}$. 
 Our interest is now in determining how the characteristics of these source and medium  
statistics determine the scintillation index of the transmitted  wave field.  We discuss next 
the equation that can be used to describe the evolution of the statistics of the wave field, that is, the equation that describes 
how the random scattering modifies the statistical distribution  of the wave field
from those of the source as the wave field  propagates through the complex medium.     

\subsection{The It\^o-Schr\"odinger Equation}
\label{sec:i2}

%We consider a   scalar wave field with  the (small) relative random fluctuations
% in the index of refraction being modeled by $\nu$  in (\ref{eq:nu}). 
The complex amplitude field  $u$ which modulates the carrier plane wave:
$$
U(z,\bx,t) = \frac{i}{2k_o} \exp(ik_oz) u(z,\bx,t)
$$
satisfies the It\^o-Schr\"odinger equation   \cite{gar09}:
\begin{equation}
\label{eq:model}
 d {u}(z,\bx,t)   =     
          \frac{ i }{2k_o} \Delta_{\bx}   {u}(z,\bx,t) dz
   +   \frac{ik_o}{2}   {u} (z,\bx ,t) \circ  d{B}(z,\bx,t) 
  , 
\end{equation}
with the initial condition in the plane $z=0$:
$$
 {u}(z= 0,\bx ,t)  = f(\bx,t) .
$$
Note that here $\Delta_\bx$ is the transverse Laplacian and $t$ is the slow time scale corresponding to the 
time  at which the source and the random medium change and that  
 it  is a frozen parameter in  (\ref{eq:model}).
This is a consequence of our assumption that the coherence times of the source and  of the 
medium are long  relative to the travel time of the field  over the range $L$. 
The derivation of Eq.~(\ref{eq:model}) from Eq.~(\ref{eq:helm}) follows the lines of the proof presented in \cite{gar09,gar14a} which deals 
with the  case of a time-independent medium.  
It is valid in the white-noise paraxial regime, when the wavelength is much smaller than the correlation radii of the source and of the medium, which are themselves much smaller than the propagation distance.
Note that in (\ref{eq:model})  the symbol $\circ$ stands for the Stratonovich stochastic integral, moreover, that 
  $B(z,\bx,t)$ is a real-valued  Brownian field over $[0,\infty) \times \RR^2 \times \RR$ with   
a covariance that derives from the model for the medium fluctuations  in (\ref{eq:nu})
 \begin{equation}
 \label{defB}
\EE[   {B}(z,\bx,t)  {B}(z',\bx',t') ] =  
\sigma_{\rm m}^2 \ell_{\rm m}
 {\min\{z, z'\}}  G\left( \frac{t-t'}{\tau_{\rm m}}\right)  {C}
 \left( \frac{\bx - \bx'}{\ell_{\rm m}}\right)   ,
\end{equation}
where $C(\boldsymbol{\chi}) = \int_\RR {\cal C}_{\rm m}(\zeta,\boldsymbol{\chi}) d\zeta$, which is such that $C({\bf 0})=1$.
The first- and second-order moments of the wave field solution of (\ref{eq:model}) has been studied in   \cite{gar09,gar09b}.
The first-order moment of the wave field is zero.
The second-order moment of the wave field (mutual coherence function) defined by
\begin{equation}
\mu_2(z,\bx,\by;\tau) = \EE\Big[ u\big(z,\bx+\frac{\by}{2},t+\tau\big) \overline{u\big(z,\bx-\frac{\by}{2},t\big)}\Big]
\end{equation}
satisfies  \cite{gar14a}
\begin{equation}
\frac{\partial \mu_2}{\partial z} = \frac{i}{k_o}  \nabla_{\bx} \cdot\nabla_{\by}  \mu_{2} + \frac{k_o^2 \sigma_{\rm m}^2 \ell_{\rm m}}{4} {U}_{2} \big(  \bx,\by; \tau\big)
 \mu_{2}   , 
\end{equation}
with the potential $U_2(\bx,\by;\tau) = G(\tau/\tau_{\rm m}) C(\by/\ell_{\rm m})-1$ and the initial condition
$\mu_2(z=0,\bx,\by;\tau)=
\EE[ f(\bx+\by/2,t+\tau)\overline{f(\bx-\by/2 ,t)}]$ that are both independent on $\bx$.
%F({\tau}/{\tau_{\rm s}})  \exp[ -|\by|^2/(4 \ell_{\rm s}^2) ]$. 
The second-order moment is given by
\begin{equation}
\mu_2(z,\bx,\by;\tau) = 
F\left(\frac{\tau}{\tau_{\rm s}}\right) 
\exp\Big[ - \frac{|\by|^2}{4\ell_{\rm s}^2}
-\frac{\sigma_{\rm m}^2 k_o^2 \ell_{\rm m} z}{4} \Big( 
1-
G\left(\frac{\tau}{\tau_{\rm m}}\right) C\left( \frac{\by}{\ell_{\rm m}}\right)   \Big)\Big]  .
\end{equation}
By inspection of the behavior of the second-order moment 
when $\tau=0$:
$$
\EE\Big[ u\big(z,\bx+\frac{\by}{2},t\big) \overline{u\big(z,\bx-\frac{\by}{2},t\big)}\Big]= 
\exp\Big[ - \frac{|\by|^2}{4\ell_{\rm s}^2}
-\frac{\sigma_{\rm m}^2 k_o^2 \ell_{\rm m} z}{4} \Big( 
1-
C\left( \frac{\by}{\ell_{\rm m}}\right)   \Big)\Big] ,
$$
we find that the scattering mean free path
$\ell_{\rm mfp}$ (that is the typical propagation distance 
over which a coherent wave becomes incoherent) is
\begin{equation}
\ell_{\rm mfp}^{-1}  = \frac{\sigma_{\rm m}^2 k_o^2 \ell_{\rm m} }{8} .
\end{equation}
Note that the scattering mean free path therefore is inversely proportional to the 
medium  correlation length $\sigma_{\rm m}^2 \ell_{\rm m} $ characterizing the strength of the scattering. 
When % the propagation distance $z \gtrsim \ell_{\rm mfp}$ and 
$C$ is smooth at zero:
$C(\boldsymbol{\chi}) = 1 -c_2 |\boldsymbol{\chi}|^2+o(|\boldsymbol{\chi}|^2)$, 
the correlation radius $\rho_{\rm c}(z)$ of the wave field  is
\begin{equation}
\rho_{\rm c}^{-2}(z) = \ell_{\rm s}^{-2} + \frac{c_2 \sigma_{\rm m}^2 k_o^2 z}{\ell_{\rm m}} .
\end{equation}
By inspection of the behavior of the second-order moment 
when $\by={\bf 0}$:
$$
\EE\Big[ u\big(z,\bx,t+\tau\big) \overline{u\big(z,\bx,t\big)}\Big]= 
F\left(\frac{\tau}{\tau_{\rm s}}\right) 
\exp\Big[ 
-\frac{\sigma_{\rm m}^2 k_o^2 \ell_{\rm m} z}{4} \Big( 
1-
G\left(\frac{\tau}{\tau_{\rm m}}\right) \Big)\Big]   ,
$$
we can see that, when
% the propagation distance $z \gtrsim \ell_{\rm mfp}$ and 
$F(s)=\exp(-s^2/4)$ and $G$ is smooth at zero,  \\
$G(s)=1-g_2 s^2+o(s^2)$, 
the coherence time $\tau_{\rm c}(z)$ of the wave field  is
\begin{equation}
\tau_{\rm c}^{-2}(z) = \tau_{\rm s}^{-2} + \frac{g_2 \sigma_{\rm m}^2 k_o^2 \ell_{\rm m} z}{\tau_{\rm m}^2} .
\end{equation}
Therefore,  for deep probing both the correlation radius  and coherence time of the wave field are proportional to the reciprocal of the 
square root of the propagation distance.
We discuss next the measurements of intensity associated with the field $u$
and the  associated scintillation index. 

\subsection{Measurements and the Challenge of Understanding Scintillation}\label{sec:i3}

The intensity at lateral location $\bx$ in the plane $z=L$ of the photodetector is
\begin{equation}
I_T(\bx) = \frac{1}{T} \int_0^T |u(L,\bx,t)|^2 dt  .
\end{equation}
The intensity profile forms a  smoothed speckle pattern. This smoothed 
speckle pattern  depends in particular  on the values of the integration time $T$,
the coherence times of the source $\tau_{\rm s}$ and of the medium $\tau_{\rm m}$ and
we  aim  to understand how. 
 
The empirical scintillation index measured by the photodetector of total aperture $A$ is
\begin{equation}
{\cal S} = \frac{
\frac{1}{|A|} \int_A I_T(\bx)^2 d\bx - \Big(  \frac{1}{|A|} \int_A I_T(\bx) d\bx \Big)^2
}
{
\Big( \frac{1}{|A|}\int_A I_T(\bx) d\bx \Big)^2
}  .
\end{equation}
A more detailed model for the detector, in particular a discussion of  
the role of finite sized pixels of  a (CCD)  camera
can be found in for instance  \cite{garniers6,svetlana}.  
If the diameter of the photodetector aperture $A$ is large (much larger than the speckle radius, i.e. the correlation radius), then
\begin{equation}
{\cal S} = \frac{\EE [ I_T({\bf 0})^2] - \EE[ I_T({\bf 0}) ]^2
}
{
 \EE[ I_T({\bf 0}) ]^2
}   ,
\end{equation}
which is equal to
\begin{equation}
{\cal S} = 
\frac{2}{T} \int_0^T \big(1-\frac{\tau}{T} \big) 
\frac{ {\rm Cov} \big( |u(L,{\bf 0},0)|^2,|u(L,{\bf 0},\tau)|^2\big) }{\EE[ |u(L,{\bf 0},0)|^2]^2} d\tau  ,
\label{eq:sciind}
\end{equation}
which is our quantity of interest. 
Note that the expectation here and below refers to expectation with respect to both
the randomness of the medium and of the source. 
Note also that it follows from  \cite{gar14a} Section 5 that   $\EE[ |u(L,{\bf 0},t)|^2]=1$.
Therefore, in order to analyze  ${\cal S}$,  it remains to compute the fourth-order moment $\EE[  |u(L,{\bf 0},0)|^2 |u(L,{\bf 0},\tau)|^2]$.
We discuss the task of computing this moment next.

\section{Fourth-order Field Moment and Scintillation}\label{sec:4}
It is convenient to introduce  the notation
\be\label{eq:ft}
 f_\tau=F\left( \frac{\tau}{\tau_{\rm s}}\right), \quad g_\tau=G\left( \frac{\tau}{\tau_{\rm m}}\right) .
\ee
We also introduce a  notation for the fourth moment 
\begin{equation}
\label{def:generalmoment}
\mu_4(z,\bx_1,\bx_2,\by_1,\by_2 ; \tau) = \EE \big[ u(z,\bx_1,t+\tau) \overline{u(z,\by_1,t+\tau)}u(z,\bx_2,t) \overline{u(z,\by_2,t)}
 \big] .
\end{equation}
Here we focus on the fourth moment,
while  in \cite{fps} moments of all orders were considered under some simplifying 
assumptions of a different type.    
 The fourth moment in (\ref{eq:sciind}) is a special case of the general fourth
moment in (\ref{def:generalmoment}) corresponding to   evaluation at one 
spatial point only.  The motivation for  introducing the general fourth moment is that we can identify a
partial  differential equation satisfied by this general moment and we will subsequently discuss
the simplification that follows from evaluating this at particular values for the arguments.  
The general fourth moment  satisfies the equation
\begin{equation}
\frac{\partial \mu_4}{\partial z} = \frac{i}{2k_o}  \Big( \Delta_{\bx_1}+ \Delta_{\bx_2}
-  \Delta_{\by_1} -  \Delta_{\by_2} \Big) \mu_{4} + \frac{k_o^2 \sigma_{\rm m}^2 \ell_{\rm m}}{4} {U}_{4} \big(  \bx_1,\bx_2,  \by_1,\by_2; \tau\big)
 \mu_{4}   , \label{eq:4}
 \end{equation}
with the generalized potential
\ba
\nonumber
&& {U}_{4}\big( \bx_1,\bx_2,  \by_1,\by_2; \tau\big)  
= 
{C}\left( \frac{\bx_1-\by_1}{\ell_{\rm m}}\right) 
+
{C}\left( \frac{\bx_2-\by_2}{\ell_{\rm m}}\right)  
+
g_\tau {C}\left( \frac{\bx_1-\by_2}{\ell_{\rm m}}\right)  
\\ && \hbox{} +
g_\tau {C}\left( \frac{\bx_2-\by_1}{\ell_{\rm m}}\right)  
-
g_\tau {C}\left( \frac{\bx_1-\bx_2}{\ell_{\rm m}}\right)  
-
g_\tau {C}\left( \frac{\by_1-\by_2}{\ell_{\rm m}}\right)  
-2 
%{C}({\bf 0}) 
 ,
\ea
and the initial condition:
$$
\mu_4(z= 0 ,\bx_1,\bx_2,\by_1,\by_2; \tau) = 
\EE \big[f(\bx_1,t+\tau) \overline{f(\by_1,t+\tau)} f(\bx_2,t) \overline{f(\by_2,t)} \big] .
$$
This follows from (\ref{eq:model}) using It\^o calculus for Hilbert space valued processes \cite{gar16a,dawson}.
Using the Gaussian property of the source and Isserlis formula, the initial condition for the fourth-order moment is:
\begin{align}
\nonumber
\mu_4(z= 0 ,\bx_1,\bx_2,\by_1,\by_2; \tau) =&
\exp\Big( - \frac{|\bx_1-\by_1|^2}{4\ell_{\rm s}^2} - \frac{|\bx_2-\by_2|^2}{4\ell_{\rm s}^2}\Big) \\
&+
f_\tau^2 
\exp\Big( - \frac{|\bx_1-\by_2|^2}{4\ell_{\rm s}^2} - \frac{|\bx_2-\by_1|^2}{4\ell_{\rm s}^2}\Big)
. \label{eq:IC} 
\end{align}

We can now express the quantity of interest, the scintillation index (\ref{eq:sciind}),  
in terms of the general fourth moment
\begin{equation}
{\cal S} = \frac{2}{T} \int_0^T \big(1-\frac{\tau}{T} \big) 
\frac{\mu_4(L,{\bf 0},{\bf 0},{\bf 0},{\bf 0}; \tau) -\mu_2(L,{\bf 0},{\bf 0} ;0)^2 }{\mu_2(L,{\bf 0},{\bf 0} ;0)^2 }
d\tau  ,
\label{eq:sciind2}
\end{equation}
where $\mu_2(L,{\bf 0},{\bf 0} ; 0)=\EE[ |u(L,{\bf 0},t)|^2]=1$.  Thus,  it is the special fourth moment 
$\mu_4(L,{\bf 0},{\bf 0},{\bf 0},{\bf 0}; \tau)$ that is needed to analyze the scintillation index. 
The explicit solution of the problem (\ref{eq:4}) is not known.  
In some scaling regimes we can, however, identify asymptotic 
solutions using the framework introduced in \cite{gar16a}. 
In Appendix \ref{secant} we discuss a  fundamental transformation of the 
fourth moment equation in (\ref{eq:4})   to a simplified problem from which the special fourth moment 
$\mu_4(L,{\bf 0},{\bf 0},{\bf 0},{\bf 0}; \tau)$ derives.

\section{ Scintillation in Canonical  Scaling Regimes}\label{sec:main}

We discuss here the three scaling  regimes for the scintillation. In these regimes we   can 
solve for the fourth moment in (\ref{eq:4}) explicitly.  This allows us to get quantitative
insight about the behavior of the scintillation and how it depends on the
characteristic parameters in the problem and we comment on this in detail. 
The first two regimes are particular cases of the far-field (or Fraunhofer) regime
$
\frac{\lambda_o L}{\min(\ell_{\rm s},\ell_{\rm m})^2} \gg 1$,
when scattering is moderate or strong
$L\gtrsim \ell_{\rm mfp}$.
The third regime is a special Fresnel regime $\frac{\lambda_o L}{\min(\ell_{\rm s},\ell_{\rm m})^2} \sim 1$ when scattering is strong $L\gg \ell_{\rm mfp}$.
The parameter determining the different regimes of scintillation is the
ratio of the  correlation radius $\ell_{\rm s}$ of the source   over the correlation radius $\ell_{\rm m}$
of the medium fluctuations.
%\be
%   \frac{\ell_{\rm s}}{\ell_{\rm m}} .
%\ee
% Then for each canonical regime the magnitude of the parameters; the strength
%of the random medium fluctuations, the correlation radius  of the random medium fluctuations
%and the total propagation distance:  
%\be
%  \sigma_{\rm m}, \quad \ell_{\rm  m},   \quad L, 
%\ee
%determine the critical context in which interesting behavior of the scintillation takes place.

\subsection{Source with Large Correlation Radius}\label{sec:res1}%
We consider first  the  regime in which the correlation radius $\ell_{\rm s}$ of the source is larger than the 
correlation radius of the medium $\ell_{\rm m}$ so that 
$\frac{\lambda_o L}{\ell_{\rm m}^2} \gg 1$ but
$ \frac{\lambda_o L}{\ell_{\rm m} \ell_{\rm s}} \lesssim 1$.
%Thus, the source fluctuates on a lateral spatial  scale that is large relative to the spatial scale at which the medium microstructure fluctuates.  
%We then identify a critical regime with interesting behavior 
%for the scintillation by assuming additionally the situation with relatively weak
%medium fluctuations and long propagation distances and we  refer this
%regime as the scintillation regime. 
We carry out the analysis of this regime in Appendix \ref{sec:proof1}
where we derive the following expression  for the scintillation index (\ref{eq:sciind2}):
 \begin{equation}
{\cal S}= \frac{2}{T} \int_0^T \big(1-\frac{\tau}{T} \big) 
\Big[ 
f_\tau^2  \exp \Big(- \frac{(1-g_\tau)\sigma_{\rm m}^2 k_o^2 \ell_{\rm m}  
 L}{2}\Big)
+
{\cal Q}_{g_\tau}(L) 
+
 f_\tau^2  {\cal Q}_{1}(L) 
\Big]
d\tau  , \label{eq:S1}
\end{equation}
where the  effective scattering kernel is given by
\begin{align}
\nonumber
{\cal Q}_g(L) =& 
\exp \Big(- \frac{\sigma_{\rm m}^2 k_o^2\ell_{\rm m} 
%C({\bf 0}) 
L}{2}\Big)   \frac{1}{2\pi}   \int_{\RR^2}\exp\Big(-\frac{ |\bs|^2}{2}\Big) \\
&\times \Big[ 
 \exp \Big( \frac{\sigma_{\rm m}^2 k_o^2 \ell_{\rm m} L g }{2} \int_0^1  C\big( \frac{L \bs s' }{k_o \ell_{\rm m} \ell_{\rm s} } \big)  d s'\Big)-1\Big]   d\bs  . \label{eq:Q}
\end{align}
When $ \frac{\lambda_o L}{\ell_{\rm m} \ell_{\rm s}} \ll 1$ the expression is simpler:
$$
{\cal Q}_g(L) = 
\exp \Big(- \frac{\sigma_{\rm m}^2 k_o^2\ell_{\rm m} L}{2}(1-g) \Big) - 
\exp \Big(- \frac{\sigma_{\rm m}^2 k_o^2\ell_{\rm m} L}{2}\Big).
$$
The kernel (\ref{eq:Q}) depends on the two-point statistics of the random medium fluctuations and reflects cumulative
scattering effects over the propagation distance $L$.   Note that $g=0$ corresponds to the situation with 
 intensities of wave fields having propagated through uncorrelated random media   and that indeed
  ${\cal Q}_0(L)=0$.   
  
 The first term in the square brackets  in (\ref{eq:S1}) corresponds to the scintillation contribution from the 
 fluctuations of  the source and this contribution  is damped by  temporal decorrelation   of the
 medium fluctuations as well as temporal averaging at the detector. 
 The second term in the square brackets is the scintillation contribution 
 produced by the random medium fluctuations and is again damped by temporal decorrelation
 of the random  medium fluctuations.  The last term in the square brackets is a cross  
 term  reflecting the scintillation contribution from the combined effect of medium 
 and source  fluctuations.

 We next discuss the behavior of the scintillation index in various special cases.  
 \begin{itemize}
 \item
 Note first that  a rapid decay of $ f_\tau$ (short coherence time) corresponds to 
 rapid decorrelation of the source. Such a rapid decorrelation
 serves  to  reduce the  scintillation index due
 to averaging by  the photodetector.
 Similarly rapid decay of $g_\tau$  corresponds  to rapid 
 decorrelation in the medium fluctuations and reduced scintillation due to averaging over
 incoherent scattering contributions. We find from (\ref{eq:S1}) that when $T$ 
 becomes much larger than the coherence times of the source and of the medium,
and assuming that $\tau\mapsto f_\tau\in L^2$ (i.e., is square-integrable) and that $g_\tau$ goes to zero at 
infinity fast enough so that $\tau \mapsto Q_{g_\tau}(L) \in L^1$ (i.e., is integrable), 
then we have
$$
{\cal S} \stackrel{T \to+\infty}{\longrightarrow}  0
$$
for any propagation distance.

%\item
%In the  strongly scattering regime when we also assume that
%\be
%  g_\tau=1 -\bar{g}\tau^2+{\cal O}(\tau^3) ,  \quad \bar{g} > 0 ,
%\ee
%we have with the averaging time $T$ at the mirror fixed
%$$
%{\cal S} \stackrel{\alpha_L >> 1}{\sim} 
% \frac{2}{T} \int_0^T \big(1-\frac{\tau}{T} \big) 
%f_\tau^2 d\tau
%\left( \frac{1 }{ 1 + \frac{\bar{\gamma}_2 L^3}{3 \ell_{\rm s}^2}  }  \right) , 
%$$
% so that the scintillation index is small for any propagation distance due to the strong 
% scattering by the medium leading to a short coherence time for the transmitted field. 
% 

\item
An interesting situation corresponds to  $g_\tau \equiv 1$, which means that the medium is frozen.
We have
$$
\frac{\mu_4(L,{\bf 0},{\bf 0},{\bf 0},{\bf 0}; \tau) - \mu_2(L,{\bf 0},{\bf 0} )^2}
{ \mu_2(L,{\bf 0},{\bf 0})^2}
= f_\tau^2    +
(1+ f_\tau^2)  {\cal Q}_{1}(L) 
.
$$
This means that, even if $\tau$ is so large that the initial fields are independent ($f_\tau=0$)   the 
intensities of the transmitted fields at the photodetector at different times are correlated, in fact the covariation degree is zero  at $L=0$, is not zero for positive $L$
and goes to zero as $L\to+\infty$.
The scintillation index   is given by
\begin{align}
\label{eq:sv}
{\cal S} =& \frac{2}{T} \int_0^T \big(1-\frac{\tau}{T} \big) 
\Big[ f_\tau^2+  (1+f_\tau^2)  {\cal Q}_1(L)  \Big]
d\tau  ,
\end{align}
so that, when $T$ becomes much larger than the coherence time of the source,
and assuming that $f_\tau\in L^2$, 
then we have
$$
%\begin{equation}
{\cal S} \stackrel{T \to+\infty}{\longrightarrow}  
{\cal Q}_1(L)  .
%\label{eq:expressscingrandrho0}
%\end{equation}
$$
This shows that  the scintillation index corresponding to averaging of the initial incoherent intensity is zero, 
while the one corresponding to the transmitted field is not.   
 \item 
 For $\tau$ smaller than both the coherence times of the source and the medium, we have
$f_\tau=g_\tau=1$ and then for $T$ similarly small
\begin{equation}
{\cal S} = \frac{\mu_4(L,{\bf 0},{\bf 0},{\bf 0},{\bf 0}; \tau) - \mu_2(L,{\bf 0},{\bf 0} ; 0)^2}
{ \mu_2(L,{\bf 0},{\bf 0} ; 0 )^2}
=1 + 2 {\cal Q}_{1}(L)  . \label{eq:ng}
\end{equation}
%The two spatial intensity distributions measured at two times separated by such a small $\tau$ are fully correlated.
Thus, initially the scintillation index is one (because the source has Gaussian distribution),  then it reaches beyond one in a mixing region
and returns  to one for large propagation distances.
  Indeed, the fluctuations of the initial field happen on a  spatial  scale that is large relative
to the scale of the field variations that are imposed by 
the random medium fluctuations  resulting in  a non-Gaussian mixture situation with  scintillation index beyond one. 
\item
Consider the strongly scattering regime so that 
\be\label{eq:strong}
\alpha_L := \frac{\sigma_{\rm m}^2 k_o^2 \ell_{\rm m} 
%C({\bf 0}) 
L}{2} \gg 1 .
\ee
Note that  then the propagation distance
is  larger than the scattering mean free path %\cite{gar09} 
since $\alpha_L = 4 L /\ell_{\rm mfp}$,
so that in the case of coherent sources 
most of the wave energy has been transferred to incoherent wave energy due to scattering.  We will in the context  of the strongly scattering regime  (\ref{eq:strong}) assume that
 $C$ is smooth and isotropic so that we have (remember $C({\bf 0})=1$):
\be
\label{eq:ex}
C(\boldsymbol{\chi}) = 
%C({\bf 0})
1 - c_2 |\boldsymbol{\chi}|^2 +o(|\boldsymbol{\chi}|^2).
\ee   
Then we   we can compute a simplified expression for $ {\cal Q}_1(L)$ and find
\ban
{\cal Q}_1(L) \stackrel{\alpha_L \gg 1}{\simeq}  \frac{1}{ 1 + \frac{c_2 \sigma_{\rm m}^2 L^3}{3 \ell_{\rm m} \ell_{\rm s}^2} }  ,
\ean
so that  ${\cal Q}_1(L)$ goes to zero when the propagation distance
 $L$ becomes large. 
\end{itemize} 

\begin{figure}[htbp] % [htbp]
\centering\includegraphics[width=6cm]{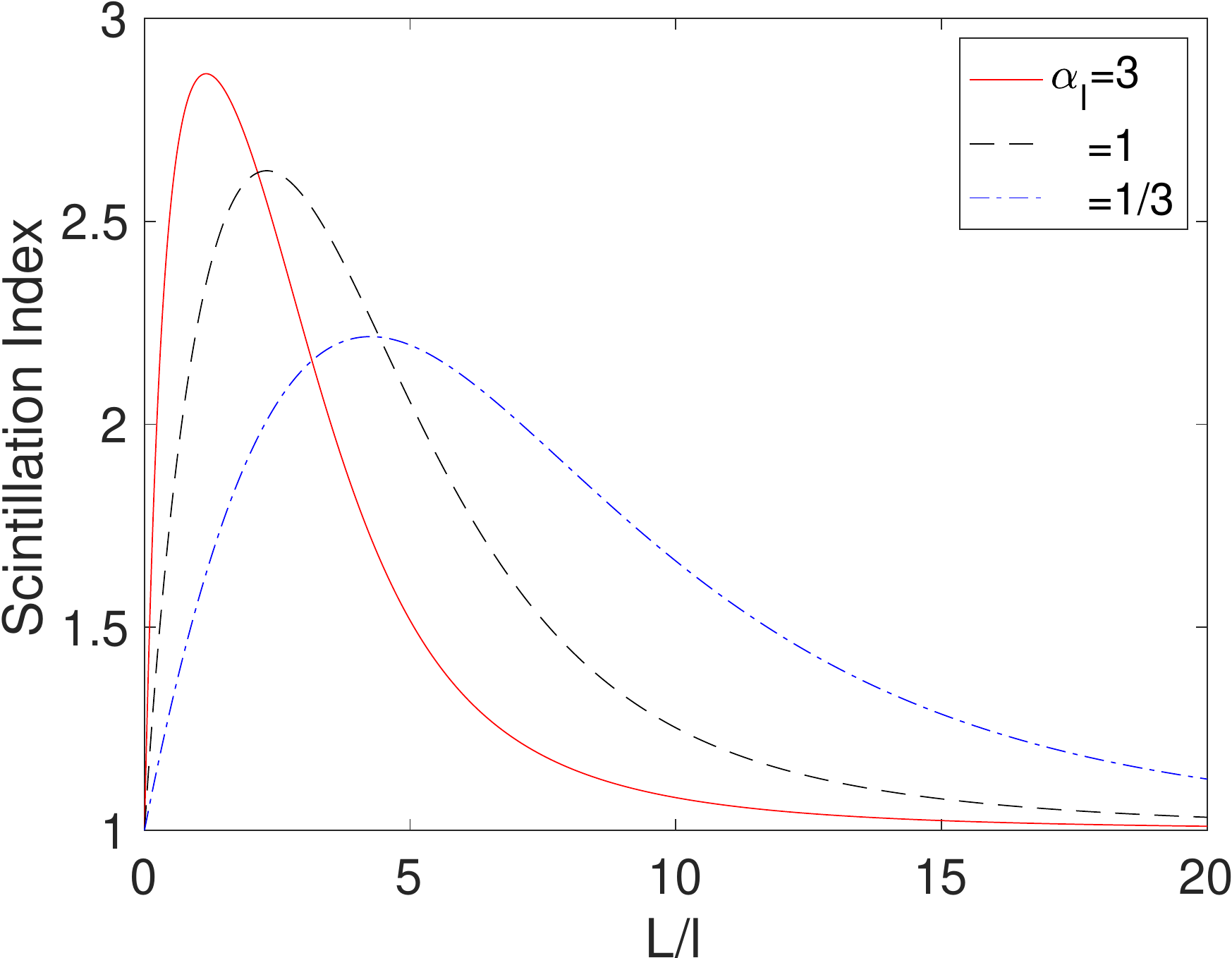}
\caption{Scintillation index as a function of the relative propagation distance $L/\ell$ for three values 
of the medium fluctuation strength parameter $\alpha_\ell$. The figure corresponds to the regime $\ell_{\rm s}\gg \ell_{\rm m}$ for small averaging times at the detector so that the scintillation index is given by (\ref{eq:ng}).   }\label{fig:4}
\end{figure}

 In Figure \ref{fig:4} we illustrate  the behavior of the scintillation in the regime $\ell_{\rm s} \gg \ell_{\rm m}$. The figure
 shows how the scintillation index in (\ref{eq:ng}) depends  on the propagation distance.
 We introduce the length parameter
 \ban
    \ell =\frac{ \ell_{\rm m} \ell_{\rm s} }{ \lambda_0  }.
 \ean 
 Then we  show the scintillation index as function of $L/\ell$ for three different values of the medium 
fluctuation strength parameter 
\ban
 \alpha_\ell  = \frac{\sigma_{\rm m}^2 k_o^2 \ell_{\rm m} 
 %C({\bf 0}) 
 \ell}{2} ,
\ean
when $C(\boldsymbol{\chi})=\exp(-|\boldsymbol{\chi}|^2/2)$. 
Note that with stronger medium fluctuations the maximum value for the scintillation index
is larger and happens for shorter propagation distances.  

\subsection{Source with Intermediate Correlation Radius}\label{sec:res2}%
We consider next  the case when the correlation radius of the source  is of the same order as 
the correlation radius of the medium
so that 
$\frac{\lambda_o L}{\ell_{\rm m}^2} \gg 1$
and 
$\frac{\lambda_o L}{\ell_{\rm m} \ell_{\rm s}} \gg 1$.
%Thus, the source fluctuates on a lateral spatial 
% scale corresponding to the spatial scale at which the medium microstructure 
% fluctuates.  
%As above, we then identify a critical regime with interesting behavior 
%for the scintillation by assuming additionally   relatively weak
%medium fluctuations and long propagation distances  and  refer 
%to this as the  scintillation regime. 
We carry out the analysis in Appendix \ref{sec:proof2}
where we derive the following expression  for the scintillation index (\ref{eq:sciind2}):
 \begin{align}\label{eq:S2} 
{\cal S} =& \frac{2}{T} \int_0^T \big(1-\frac{\tau}{T} \big)  
\Big[ f_\tau^2  \exp \Big(- \frac{(1-g_\tau) \sigma_{\rm m}^2 k_o^2  \ell_{\rm m} 
L }{2}  \Big)\Big] d\tau .
\end{align}
As above the term in the square brackets   corresponds to scintillation contribution from the 
 fluctuations in the source and this contribution  is damped by  fast temporal decorrelation   of the
 medium fluctuations (small $g_\tau$), moreover, by smoothing at the detector. 
 Note that this term corresponds to the first term in the square brackets in (\ref{eq:S1}). 
 Indeed, the last two terms in the square brackets in (\ref{eq:S1}) becomes small when 
 $\ell_{\rm s} $ is reduced so that $\frac{\lambda_o L}{\ell_{\rm m}\ell_{\rm s}}\gg1$.  The situation with $\ell_{\rm s} \sim \ell_{\rm m}$ 
 is the regime considered here. If we further reduce $\ell_{\rm s}$ so that
 $\ell_{\rm s} \ll \ell_{\rm m}$ then we may transition towards the regime considered in 
 the next section.  
  In the regime $\ell_{\rm s} \sim \ell_{\rm m}$ we can observe the following behaviors.
\begin{itemize}
\item
In the case when $T$ becomes much larger than the coherence time of the source,
and assuming that $f_\tau\in L^2$, 
we have
$$
{\cal S} \stackrel{T \to+\infty}{\longrightarrow}  0
$$
for any propagation distance due to averaging at the photodetector. 
  \item
For $\tau$ smaller than the coherence times of both the source and the medium, we have
$f_\tau=g_\tau=1$ and then  with $T$ similarly small
$$
{\cal S} =1    .
$$
Here the field behaves as a complex Gaussian field, because the fluctuations
in the source and in the medium happen at the same scale.
\end{itemize} 
    
In  Figures  \ref{fig:5} and \ref{fig:6} we illustrate  the behavior of the scintillation  index in the regime $\ell_{\rm s}\sim \ell_{\rm m}$. The figures
 show how the scintillation index depends 
 on the  magnitudes of the coherence times of the source 
 and of the medium relative to the integration time 
 at the detector. The two figures correspond to two different values  
 for the strength of the medium fluctuations parameter $\alpha_L$. 
 The time coherence functions $f_\tau$ and $g_\tau$ are chosen to be Gaussian
 \begin{equation}
 \label{eq:exftau}
    f_\tau = \exp\left( - \frac{\tau^2}{2 \tau_{\rm s}^2} \right), \quad  
    g_\tau = \exp\left( - \frac{\tau^2}{2 \tau_{\rm m}^2} \right) . 
 \end{equation}
    Note how strong medium fluctuations serve to reduce the scintillation index in the case with a time-dependent 
 random medium
 and temporal averaging at the detector, moreover, how  also long duration detector temporal averaging serves to reduce 
 the  scintillation index. 
 \begin{figure}
\centerline{
\begin{picture}(226,185)
\put(20,8)
{\includegraphics[width=0.5\linewidth]{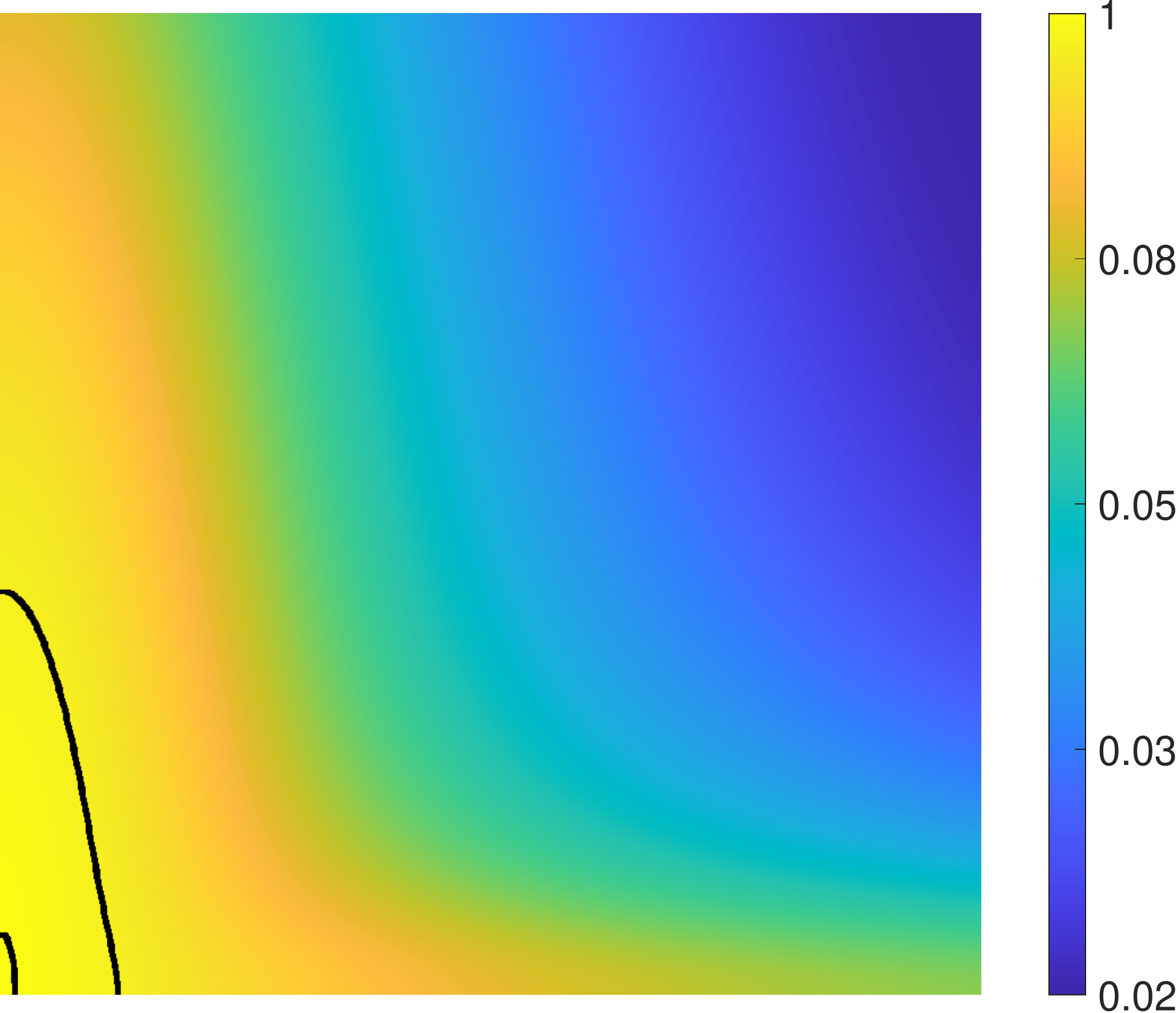}}  
\put(20,10){\vector(1,0){167}}
\put(20,10){\vector(0,1){170}}
\put(15,2){$0$}
 \put(-27,167){$T/\tau_{\rm s}=20$}
  \put(160,-2){$T/\tau_{\rm m}=80$}
     \end{picture}
}  
\caption{Scintillation index ${\cal S}$ as a function of the coherence times of the 
source  and of the random medium, respectively $\tau_{\rm s}$ and $\tau_{\rm m}$, relative to $T$,  the averaging 
time  at the detector, in the regime $\ell_{\rm s} \sim \ell_{\rm m}$. The effective strength parameter for the magnitude of the medium fluctuations is 
$\alpha_L=\sigma_{\rm m}^2 k_o^2 \ell_{\rm m}
% C({\bf 0})
L/2=3$.  The two solid black lines correspond  to the contour levels ${\cal S}=.8$ and $.2$ respectively.
Note that here and below we adapt the color scale to the particular distribution of scintillation index values.  
}
 \label{fig:5}
\end{figure}
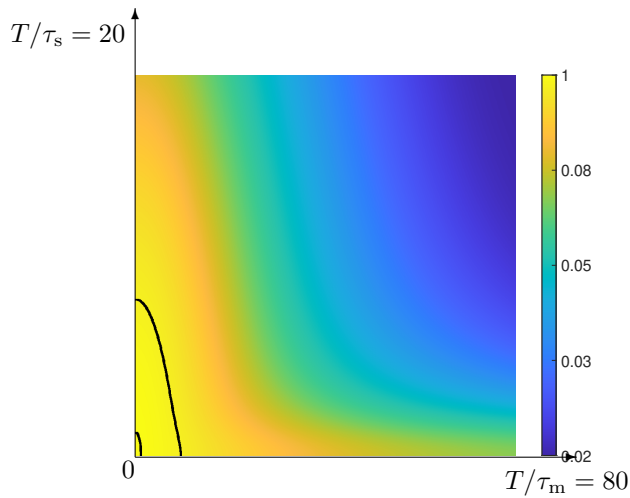  
 \begin{figure}
\centerline{
\begin{picture}(226,185)
\put(20,9)
{\includegraphics[width=0.5\linewidth]{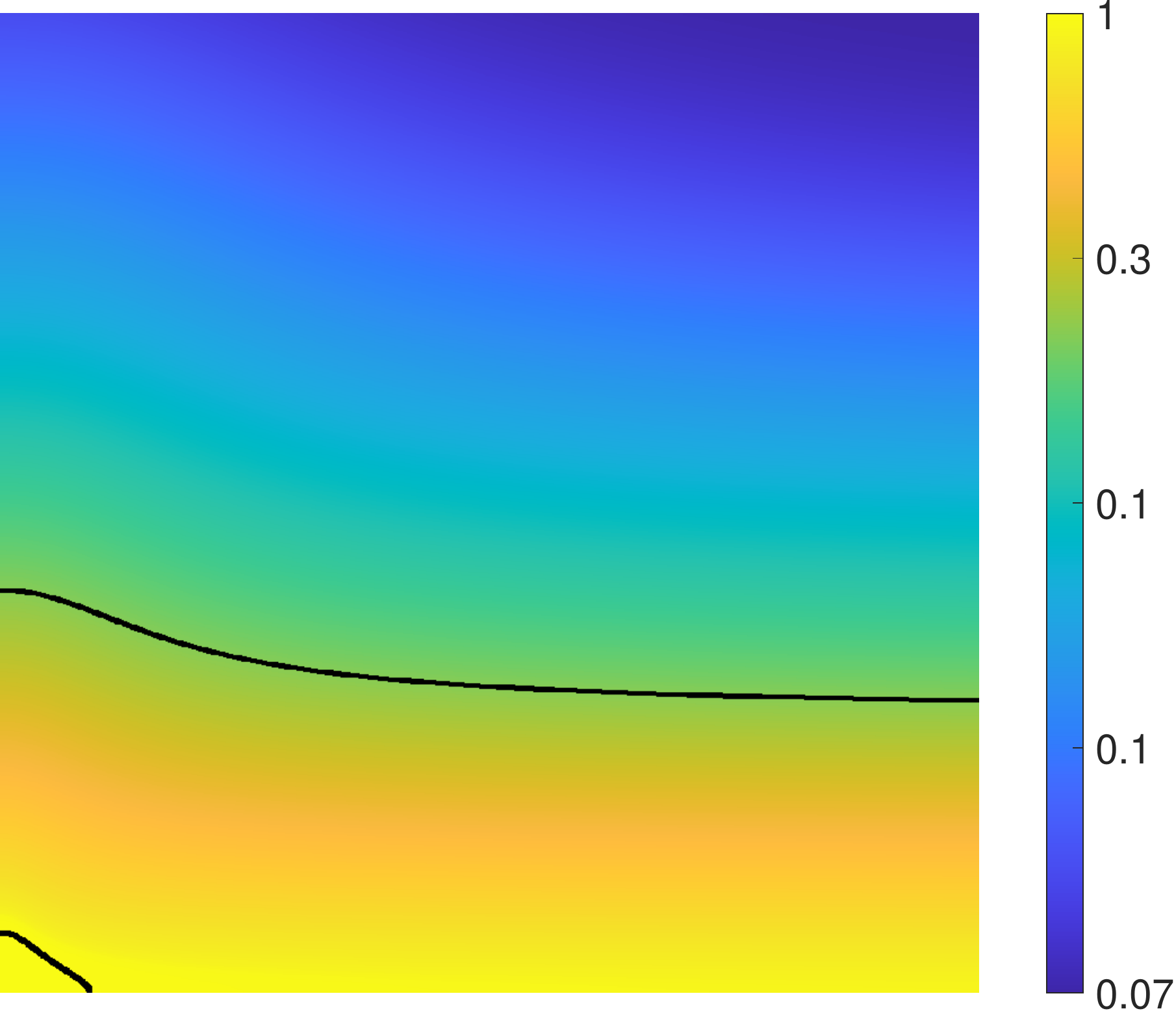}}  
\put(20,10){\vector(1,0){167}}
\put(20,10){\vector(0,1){170}}
\put(15,2){$0$}
 \put(-27,167){$T/\tau_{\rm s}=20$}
  \put(160,-2){$T/\tau_{\rm m}=80$}
     \end{picture}
}  
\caption{As in Figure \ref{fig:5},  however with the medium fluctuation strength parameter
$\alpha_L= \sigma_{\rm m}^2 k_o^2 
%C({\bf 0})
\ell_{\rm m} L/2=1/3$. 
}
 \label{fig:6}
\end{figure}
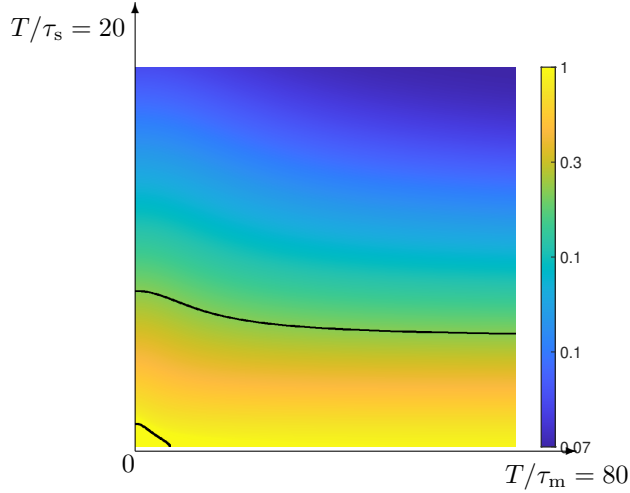

\subsection{Source with Small Correlation Radius}\label{sec:res3}%
We finally consider the regime in which the correlation radius of the source is smaller  than the correlation radius of the medium and we have 
$\frac{\lambda_o L}{\ell_{\rm s}^2} \lesssim 1$.
A similar regime (called spot-dancing regime) has already been considered in the literature to study coherent and narrow beam propagation: 
the beam propagates with the same transverse profile as in a homogeneous medium but its center randomly wanders, 
more exactly,  its center is a random  process whose standard deviation increases with propagation distance  \cite{gar14a,furutsu73}.
%When the soure has small correlation radius we consider a rescaling where the random fluctuations are relatively
%strong and decorrelate relatively slowly, it is in this regime that the scintillation shows an  interesting behavior. 
Here, we also assume that ${C}$ is smooth and isotropic with an expansion as 
in (\ref{eq:ex}). 
We carry out the analysis in Appendix \ref{sec:proof3}
 where we derive the following expression for the scintillation index (\ref{eq:sciind2}):
 \begin{align}
{\cal S} =& \frac{2}{T} \int_0^T \big(1-\frac{\tau}{T} \big)  
\bigg[ 
\frac{ f_\tau^2 }{1+ 
\frac{ (1-g_\tau)  c_2 \sigma_{\rm m}^2 L^3}{3 \ell_{\rm m} \ell_{\rm s}^2}} 
\bigg]  
d\tau  .  \label{eq:S3} 
\end{align}
The numerator within the square brackets corresponds to the scintillation  contribution of the Gaussian source,
while the denominator  corresponds to damping of  the scintillation index due to temporal decorrelation
in the random medium.  
We can moreover make the following observations.
\begin{itemize}
\item
When $T$ becomes much larger than the coherence time of the source,
and assuming that $f_\tau\in L^2$,  then
we have
$$
{\cal S} \stackrel{T \to+\infty}{\longrightarrow} 0
$$
for any propagation distance. Note that the scintillation index is small even if the  medium is frozen
since the medium fluctuations do not strongly contribute to the intensity 
correlations with the very small source  correlation radius.

\item
For $\tau$ smaller than the coherence times of the source  and of the medium, we have
$f_\tau=g_\tau=1$ and it follows that with $T$ similarly small 
$$
%\begin{equation}
{\cal S} 
= 1  .
%\end{equation}
$$
This  corresponds to a Gaussian situation since the random medium 
fluctuations again  does not  strongly affect the correlations  in this case 
with a source with rapid stationary spatial fluctuations.  
This is in contrast to the situation with
a deterministic beam source when the spot dancing property produces a  heavy-tailed intensity distribution  and large scintillation index
(a non-central chi-square distribution with two degrees of freedom, also known as the Rice-Nakagami distribution
\cite{gar14a}).
 \end{itemize} 
 
 In {Figures}   \ref{fig:7} and \ref{fig:8} we illustrate  the behavior of the scintillation index in the regime $\ell_{\rm s} \ll \ell_{\rm m}$. The figure
 shows how the scintillation index depends 
 on the  magnitudes of the coherence times of the source 
 and of the medium relative to the integration time 
 at the detector. The two figures correspond to two different values  
 for the effective  strength of the medium fluctuations 
  \ban
  \frac{c_2 \sigma_{\rm m}^2 L^3}{3\ell_{\rm m} \ell_{\rm s}^2}   .
 \ean
 and    
  we again assume a Gaussian time coherence function   $f_\tau$ as in (\ref{eq:exftau}).   Note that in this case the strength parameter does not depend on  the central wavelength $\lambda_0$. 
  As above note how strong medium fluctuations serve to reduce the scintillation index in the case 
  with a time-dependent  random medium
 and temporal averaging at the detector, moreover, how  again long detector temporal averaging serves to reduce 
 the  scintillation index. 
 \begin{figure}
\centerline{
\begin{picture}(226,185)
\put(20,8)
{\includegraphics[width=0.5\linewidth]{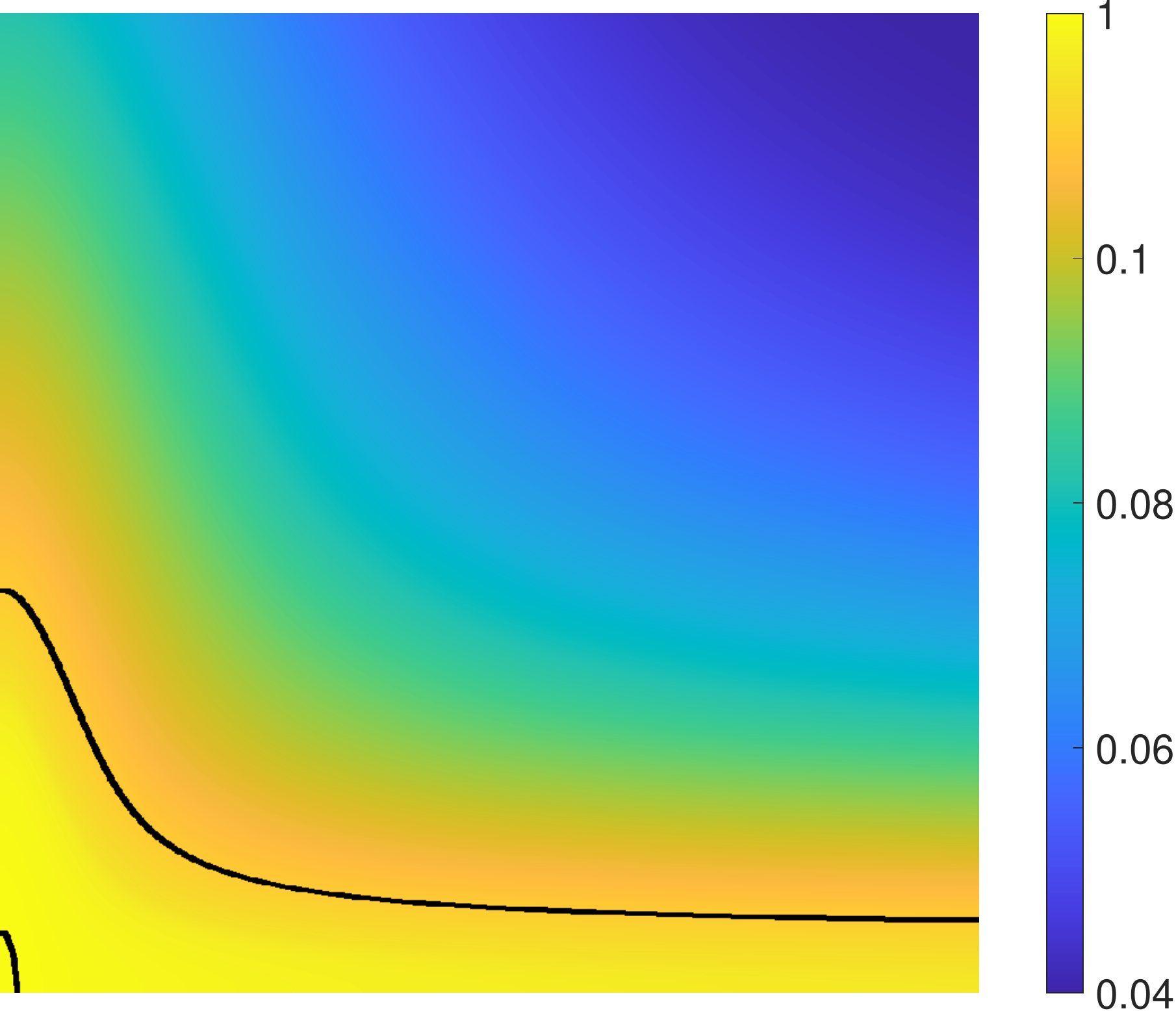}}  
\put(20,10){\vector(1,0){167}}
\put(20,10){\vector(0,1){170}}
\put(15,2){$0$}
 \put(-27,167){$T/\tau_{\rm s}=20$}
  \put(160,-2){$T/\tau_{\rm m}=80$}
     \end{picture}
}  
\caption{Scintillation index ${\cal S}$ as a function of the coherence times of the 
source  and of the random medium, respectively $\tau_{\rm s}$ and $\tau_{\rm m}$, relative to $T$,  the averaging 
time  at the detector, in the regime $\ell_{\rm s} \ll \ell_{\rm m}$. The effective strength parameter for the magnitude of the medium fluctuations is
$c_2 \sigma_{\rm m}^2 L^3/(3 \ell_{\rm m} \ell_{\rm s}^2)=3$.  The two solid black lines correspond  to the contour levels ${\cal S}=.8$ and $.2$ respectively. 
}
 \label{fig:7}
\end{figure}
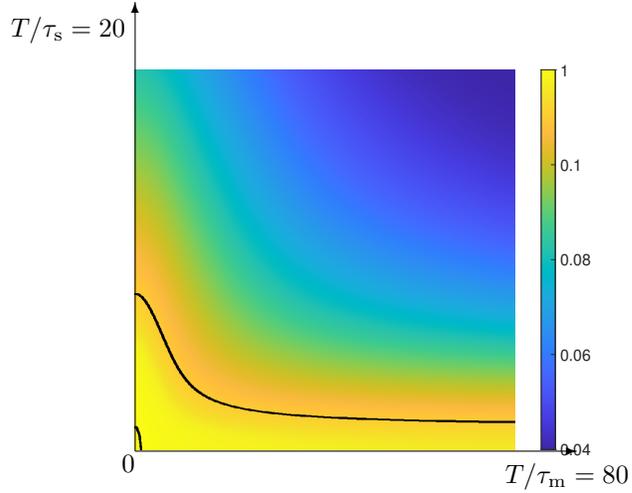  
 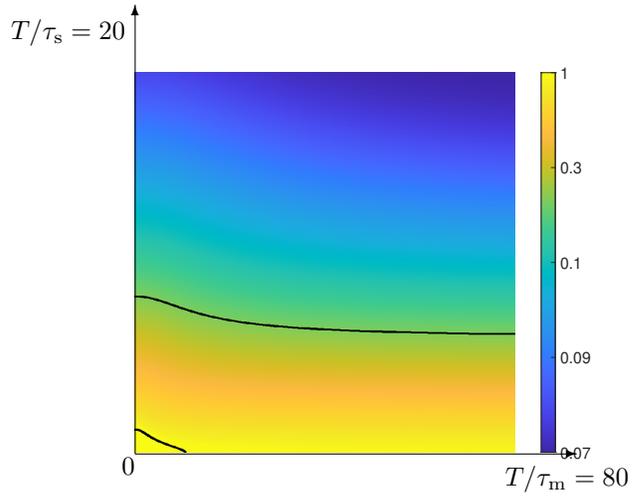
\begin{figure}
\centerline{
\begin{picture}(226,185)
\put(20,8)
{\includegraphics[width=0.5\linewidth]{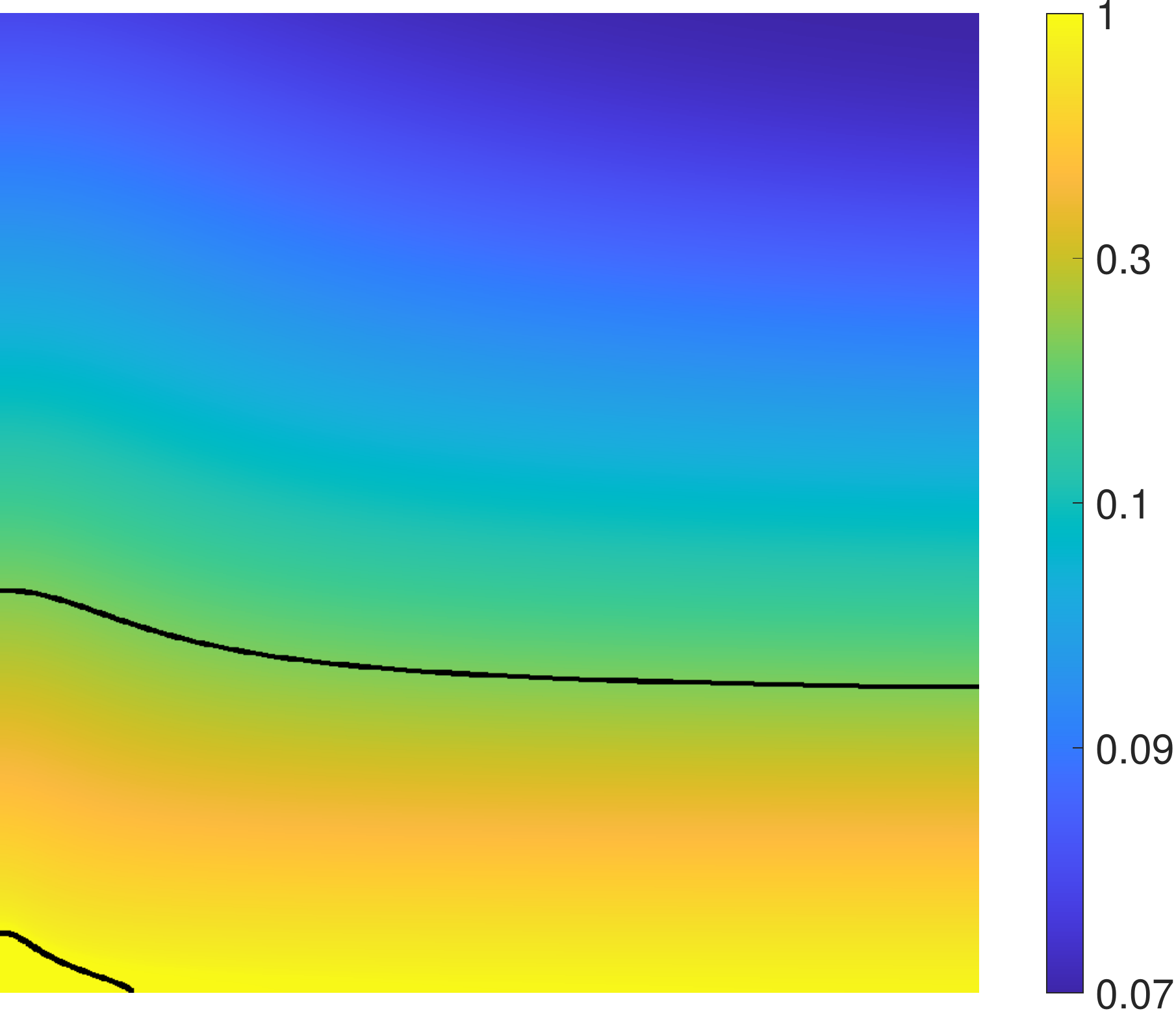}}  
\put(20,10){\vector(1,0){167}}
\put(20,10){\vector(0,1){170}}
\put(15,2){$0$}
 \put(-27,167){$T/\tau_{\rm s}=20$}
  \put(160,-2){$T/\tau_{\rm m}=80$}
     \end{picture}
}  
\caption{As in Figure \ref{fig:7},  however with
medium fluctuation strength parameter
$c_2 \sigma_{\rm m}^2 L^3/(3 \ell_{\rm m} \ell_{\rm s}^2)=1/3$. 
}
 \label{fig:8}
\end{figure}

\section{Example with Experimental Data}\label{sec:S} 

\begin{figure}[h!] % [htbp]
\centering\includegraphics[width=6cm]{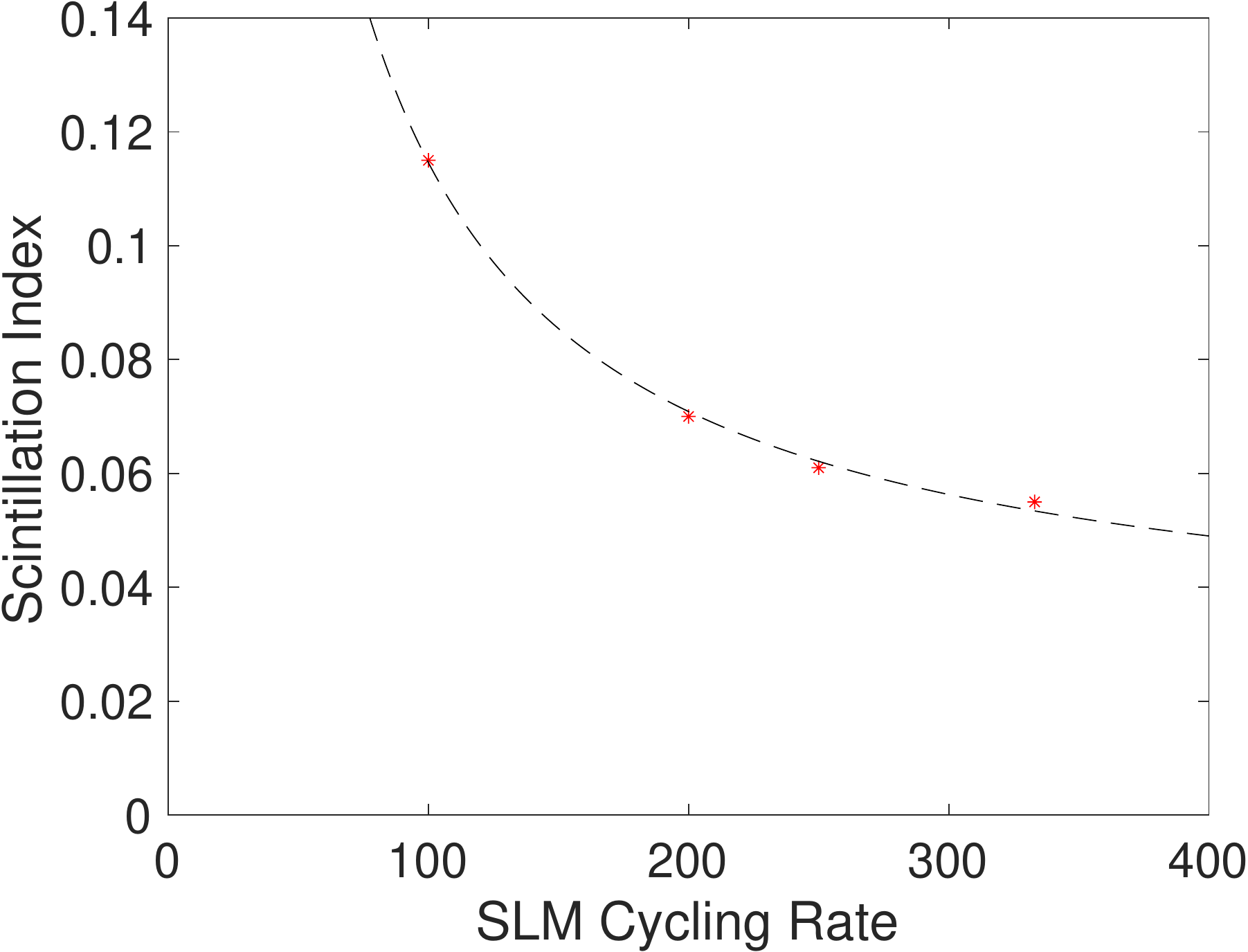}
\caption{Measurements  of scintillation index as function of the SLM cycling rate (red stars).
The observations conform well with the theoretical predictions (dashed line) assuming a frozen medium. }\label{fig:9}
\end{figure}

We discuss an example with real data taken from the paper \cite{svetlana}
by Nelson et al. 
 The experiment in \cite{svetlana} involves  an 
 over-the-water %(College Creek) 
 laser beam link at the United States Naval Academy.
 The  source is partially coherent (Multi-Gaussian Schell Model) and realized via a SLM.
 The measurement procedure at the CDD camera  corresponds to an averaging interval
 of $T=60${\it sec}. The experiment is carried out for  various values for the source coherence  time $\tau_{\rm s}$   
 realized via varying the SLM cycling rate.  The field trials were conducted in July and were performed 
 during the night in calm weather conditions over a maritime link of 323 meters.
We refer to the paper \cite{svetlana}  for a more detailed description of the experimental setup.
Assuming a frozen medium in view of the calm weather
 we can then model the observed scintillation index as
in (\ref{eq:sv}) 
 \begin{equation}
{\cal S} =\frac{2}{T} \int_0^T \big(1-\frac{\tau}{T} \big) 
\Big[ f_\tau^2+  (1+f_\tau^2)  {\cal Q}_1(L)  \Big]
d\tau      \stackrel{\tau_{\rm s} \ll T }{\longrightarrow}  \frac{c_1}{\tau_{\rm s}^{-1}} + c_2 ,
\end{equation}
and we can fit  the parameters $c_1,c_2$  via least squares. The results are shown in Figure \ref{fig:9}
and we can see an excellent fit in between model and data.    

\section{Conclusions} 

 We have considered the scintillation of wave field that is observed  after propagation  through
 a time-dependent random medium. The source is partially coherent in time and space and  constitutes
 a random field in lateral space and time variables.    We consider a high-frequency and far-field regime.  
 We give here precise characterizations of the scaling regimes leading to the different
  canonical forms of scintillation.  
  The central scaling parameters are the temporal and spatial 
 statistical coherence lengths  of the source and of the random medium, 
 in addition to the propagation range and the strength of the random medium fluctuations, and the response time of the photodetector.
 In the high-frequency and far-field regime, 
three scaling regimes
  are identified   depending
 on the magnitude of the spatial correlation radius  of  the source relative to that of the medium. 
  We identify general formulas for the scintillation index in each regime and discuss
  special cases corresponding to an effective  Gaussian situation with scintillation index
  being equal to one, a non-Gaussian mixture situation with scintillation index reaching beyond one,
  and situations with small scintillation index corresponding to a desirable high signal-to-noise
  ratio  for the measured intensity. In particular temporal averaging creates situations
  with a low  scintillation index.   
     In the context
  of for instance communication, however,  long averaging times are not in general desirable 
  and   our analysis presents quantitative insights about appropriate tradeoffs 
  that can  be made for optimal system performance. Such particular system optimization 
  challenges are left for future work.     
   We also remark that  we have  considered the case
  when the source has infinite lateral spatial extent and is a stationary stochastic process
  in  lateral  space coordinates and time. The case when the source is modulated by a finite source
  aperture and the associated challenge of identifying the spreading of the wave field  and the 
  evolution
  of speckle statistics can be analyzed via   similar theoretical frameworks as those presented here, 
  but is also left for future work. 
  
\appendix

\section{Analysis of Fourth-order Moment Equations}\label{secant}
 The main equation underlying the above results is a simplified
 equation  deriving from (\ref{eq:4}) and from which  the expressions of the special  fourth moments $\mu_4(L,{\bf 0},{\bf 0},{\bf 0},{\bf 0}; \tau)$
 follow.   
 We deduce this equation  here and analyze it in the specific scintillation regimes  in  Appendix \ref{app:2}.  
 
 Consider the general moments in (\ref{def:generalmoment}) satisfying (\ref{eq:4}) with initial condition (\ref{eq:IC}).  
It will be convenient to  parameterize  the four points 
$\bx_1,\bx_2,\by_1,\by_2$ in (\ref{def:generalmoment}) in the special way:
\begin{eqnarray}
\label{eq:reliexr1}
\bx_1 = \frac{\br_1+\br_2+\bq_1+\bq_2}{2}, \quad \quad 
\by_1 = \frac{\br_1+\br_2-\bq_1-\bq_2}{2}, \\
\bx_2 = \frac{\br_1-\br_2+\bq_1-\bq_2}{2}, \quad \quad 
\by_2 = \frac{\br_1-\br_2-\bq_1+\bq_2}{2}.
\label{eq:reliexr2}
\end{eqnarray}
In particular $\br_1/2$ is the barycenter of the four points $\bx_1,\bx_2,\by_1,\by_2$:
\begin{eqnarray*}
\br_1 = \frac{\bx_1+\bx_2+\by_1+\by_2}{2} , \quad \quad 
\bq_1 = \frac{\bx_1+\bx_2-\by_1-\by_2}{2}, \\
\br_2 = \frac{\bx_1-\bx_2+\by_1-\by_2}{2}, \quad \quad 
\bq_2 = \frac{\bx_1-\bx_2-\by_1+\by_2}{2}.
\end{eqnarray*}
We denote by $\mu$ the fourth-order moment in these new variables:
\begin{equation}
\mu(z,\bq_1,\bq_2,\br_1,\br_2; \tau) = 
\mu_4  (z, 
\bx_1  ,
\bx_2 ,
\by_1 ,  
\by_2 ; \tau
)
\end{equation}
with $\bx_1,\bx_2,\by_1,\by_2$ given by (\ref{eq:reliexr1}-\ref{eq:reliexr2}) in terms of $\bq_1,\bq_2,\br_1,\br_2$.

In the variables $(\bq_1,\bq_2,\br_1,\br_2)$ the function ${\mu} $ satisfies the system:
\begin{equation}
\label{eq:M20}
\frac{\partial {\mu}}{\partial z} = \frac{i}{k_o} \big( \nabla_{\br_1}\cdot \nabla_{\bq_1}
+
 \nabla_{\br_2}\cdot \nabla_{\bq_2}
\big)  {\mu} + \frac{\sigma_{\rm m}^2 k_o^2 \ell_{\rm m}}{4} {U}(\bq_1,\bq_2,\br_1,\br_2; \tau) {\mu}   ,
\end{equation}
with the generalized potential
\ba && 
\nonumber
{U}(\bq_1,\bq_2,\br_1,\br_2; \tau) =
{C}\left( \frac{\bq_2+\bq_1}{\ell_{\rm m}}\right) 
+
{C}\left( \frac{\bq_2-\bq_1}{\ell_{\rm m}}\right) 
+
g_\tau {C}\left( \frac{\br_2+\bq_1}{\ell_{\rm m}}\right) 
\\ && \hbox{}  +
g_\tau 
{C}\left( \frac{\br_2-\bq_1}{\ell_{\rm m}}\right) - 
g_\tau {C}\left( \frac{ \bq_2+\br_2}{\ell_{\rm m}}\right) - g_\tau {C}\left( \frac{ \bq_2-\br_2}{\ell_{\rm m}}\right) - 2
% {C}({\bf 0}) 
 .
\ea
The Fourier transform (in $\bq_1$, $\bq_2$, $\br_1$, and $\br_2$) of the fourth-order moment
is defined by:
\begin{align}
\nonumber
\hat{\mu}(z,\bxi_1,\bxi_2,\bzeta_1,\bzeta_2; \tau) 
=& 
\iint_{\RR^2 \times \RR^2 \times \RR^2 \times \RR^2} {\mu}(z,\bq_1,\bq_2,\br_1,\br_2; \tau)  \\
& \hspace*{-0.8in}
\times
\exp  \big(- i\bq_1 \cdot \bxi_1- i\bq_2 \cdot \bxi_2- i\br_1\cdot \bzeta_1- i\br_2\cdot \bzeta_2\big) d\bq_1d\bq_2 
d\br_1d\br_2 \label{eq:fourier} 
. \hspace*{0.3in} 
\end{align}
It satisfies
\begin{eqnarray}
\nonumber
&& 
\frac{\partial \hat{\mu}}{\partial z} + \frac{i}{k_o} \big( \bxi_1\cdot \bzeta_1+   \bxi_2\cdot \bzeta_2\big) \hat{\mu}
=
\frac{\sigma_{\rm m}^2 k_o^2 \ell_{\rm m}^3}{4 (2\pi)^2} 
\int_{\RR^2} \hat{C}(\bk\ell_{\rm m}) \bigg[  
 \hat{\mu} (  \bxi_1-\bk, \bxi_2-\bk, \bzeta_1, \bzeta_2)  \\
\nonumber
&& \quad   + \hat{\mu} (  \bxi_1+\bk, \bxi_2-\bk, \bzeta_1, \bzeta_2)
  -
2 \hat{\mu}(\bxi_1,\bxi_2, \bzeta_1, \bzeta_2)    \\
\nonumber
&&  \quad 
+g_\tau  \hat{\mu} (  \bxi_1+\bk,\bxi_2, \bzeta_1,  \bzeta_2-\bk)  + 
g_\tau  \hat{\mu} (  \bxi_1-\bk,\bxi_2,  \bzeta_1, \bzeta_2-\bk)    
\\
&&  \quad 
-g_\tau  \hat{\mu} (  \bxi_1,\bxi_2-\bk, \bzeta_1, \bzeta_2-\bk)  
-g_\tau \hat{\mu} (  \bxi_1,\bxi_2+\bk,  \bzeta_1, \bzeta_2-\bk) 
\bigg]d \bk ,
\label{eq:fouriermom0}
\end{eqnarray}
starting from 
\begin{align}
\nonumber
\hat{\mu}(z=0,\bxi_1,\bxi_2,\bzeta_1,\bzeta_2; \tau) =& 
(2\pi)^8  \phi_{\ell_{\rm s}^{-1}}(\bxi_1)  \phi_{\ell_{\rm s}^{-1}}(\bxi_2) \delta(\bzeta_1) \delta(\bzeta_2)
\\
& +
(2\pi)^8   f_\tau^2 \phi_{\ell_{\rm s}^{-1}}(\bxi_1)  \phi_{\ell_{\rm s}^{-1}}(\bzeta_2) \delta(\bzeta_1) \delta(\bxi_2)  .
\end{align}
Here $\hat{C}(\bq) = \int_{\RR^2} C(\boldsymbol{\chi})
\exp(i \bq \cdot \boldsymbol{\chi})
d\boldsymbol{\chi}$ is the Fourier transform of $C$,
\begin{equation}
\phi_{\kappa}(\bxi) = \frac{1}{2\pi \kappa^2 } \exp \Big( - \frac{ |\bxi|^2}{2\kappa^2}\Big)  ,
\end{equation}
is the $2$-dimensional centered isotropic Gaussian density with standard deviation 
$\kappa$, $\delta$ is the Dirac delta distribution, and $f_\tau, g_\tau $ are defined in (\ref{eq:ft}). 
We now seek to characterize 
\begin{align}
\nonumber
\mu_4(L,{\bf 0},{\bf 0},{\bf 0},{\bf 0}; \tau) =& \mu(L,{\bf 0},{\bf 0},{\bf 0},{\bf 0}; \tau) \\
=& \frac{1}{(2\pi)^8}
\iint_{\RR^2 \times \RR^2 \times \RR^2 \times \RR^2} \hat{\mu}(L,\bxi_1,\bxi_2,\bzeta_1,\bzeta_2; \tau) d\bxi_1 d\bxi_2 d\bzeta_1 d\bzeta_2 .
\end{align}
Thanks to the special initial condition that is proportional to $\delta(\bzeta_1)$,
the solution $\hat{\mu}$ to (\ref{eq:fouriermom0}) is itself proportional to $\delta(\bzeta_1)$,
and we can therefore reduce the problem (\ref{eq:fouriermom0}) to the analysis of
\begin{equation}
\hat{\eta} (z,\bxi_2,\bzeta_2; \tau) = 
\frac{1}{(2\pi)^4} \iint_{\RR^2 \times \RR^2} \hat{\mu}(z,\bxi_1,\bxi_2,\bzeta_1,\bzeta_2; \tau) d\bxi_1  d\bzeta_1 .
\label{eq:defhateta}
\end{equation}
The quantity of interest is then 
\begin{equation}
\label{eq:expressmu4eta}
\mu_4(L,{\bf 0},{\bf 0},{\bf 0},{\bf 0}; \tau) =
\frac{1}{(2\pi)^4} 
\iint_{\RR^2 \times \RR^2} \hat{\eta}(L , \bxi_2 ,\bzeta_2; \tau)   d\bxi_2   d\bzeta_2  .
\end{equation}
The function $\hat{\eta}(z,\bxi_2,\bzeta_2)$ is solution of  the characteristic system 
\begin{eqnarray}
\nonumber
&& 
\frac{\partial \hat{\eta}}{\partial z} + \frac{i}{k_o}    \bxi_2\cdot \bzeta_2  \hat{\eta}
=
\frac{\sigma_{\rm m}^2 k_o^2 \ell_{\rm m}^3 }{4 (2\pi)^2} 
\int_{\RR^2} \hat{C}(\bk\ell_{\rm m}) \bigg[    -
2 \hat{\eta}( \bxi_2, \bzeta_2) +
2 \hat{\eta} (   \bxi_2-\bk,   \bzeta_2)    \\
&&  + 
2g_\tau  \hat{\eta} (  \bxi_2,   \bzeta_2-\bk)    
-g_\tau  \hat{\eta} (   \bxi_2-\bk,  \bzeta_2-\bk)  
-g_\tau \hat{\eta} (  \bxi_2+\bk , \bzeta_2-\bk) 
\bigg]d \bk ,
\label{eq:fouriermom0eta}
\end{eqnarray}
starting from 
\begin{equation}\label{eq:4ic}
\hat{\eta}(z=0, \bxi_2 ,\bzeta_2; \tau) =
(2\pi)^4  \phi_{\ell_{\rm s}^{-1}}(\bxi_2)  \delta(\bzeta_2) +
(2\pi)^4   f_\tau^2   \phi_{\ell_{\rm s}^{-1}}(\bzeta_2)   \delta(\bxi_2)  .
\end{equation}
 This simplified system (\ref{eq:fouriermom0eta}-\ref{eq:4ic}) underlies  the scintillation results  presented above.
 Note that the fourth moment problem has been reduced to a problem defined 
 relative to two, rather than four, copies of the lateral spatial variables.  
We derive explicit solutions of this  system 
 in different scaling regimes in the 
 %next section.
next Appendix \ref{app:2}.

\section{Derivation of Scintillation Results}
\label{app:2}%
As mentioned in Section \ref{sec:i2},
the It\^o-Schr\"odinger equation is valid in the white-noise paraxial regime, when the wavelength is much smaller than the correlation radii of the source and of the medium, which are themselves much smaller than the propagation distance.
By the It\^o-Schr\"odinger equation, the fourth-order moment (\ref{def:generalmoment}) satisfies a closed equation  (\ref{eq:4}).
In this section, we derive closed form expressions of the solution of Eq.~(\ref{eq:4}) in three special white-noise paraxial  regimes, depending on the ratio of the correlation radii of the source and of the medium.

\subsection{Scintillation Regime with a Large Correlation Radius of the Source}\label{sec:proof1}
We consider the white-noise paraxial regime in which, additionally, the correlation radius of the source is larger than 
the correlation radius of the medium $\ell_{\rm s}\gg \ell_{\rm m}$ and derive the results presented in Section
\ref{sec:res1}. 
More exactly, we here deal with the following scaled regime:
\begin{equation}
\label{eq:r1}
    \frac{\ell_{\rm m}}{\ell_{\rm s}} \sim  \eps  \, ,  \quad\quad
     \frac{L}{\ell_{\rm s}} \sim   \alpha^{-1}  \, ,   \quad\quad
     \frac{\lambda_o}{\ell_{\rm s}} \sim \alpha \eps   \,   ,    \quad\quad
     \sigma_{\rm m}^2 \sim \alpha^3 \eps \,      ,
\end{equation}
and we assume $  \alpha \ll \eps \ll 1$
[Note that $L/ \ell_{\rm mfp} \sim 1$ and $\lambda_o L /\ell_{\rm m}^2 \sim \eps^{-1}$].
This means that the paraxial white-noise limit $\alpha \to 0$ is taken first (and we get an $\eps$-dependent It\^o-Schr\"odinger equation), and then we want to apply the limit  $\eps\to 0$ in the fourth-moment equation (\ref{eq:fouriermom0eta}).
%and we find
%$$ 
%2ik_o d {u}^\eps      
%    +\Delta_{\bx}   {u}^\eps \, dz
%   +  k_o^2   {u}^\eps   \circ  d{B}^\eps(z,\bx) 
% =0 ,
%$$
%where the Brownian field ${B}^\eps$ has covariance $C^\eps$.
%Then we want to apply the limit  $\eps\to 0$.
%We introduce a small dimensionless parameter $\eps$ and we rescale
%the problem via the replacements:  
% \begin{equation}
% \label{eq:r1}
%\ell_{\rm s} \to \frac{\ell_{\rm s}}{\eps},\quad C(\bx) \to \eps C(\bx), \quad L \to \frac{L}{\eps}.
%\end{equation}
In view of (\ref{eq:r1}) it is natural to introduce the rescaled function
\begin{equation}
\label{eq:renormhatM2}
\tilde{\eta}^\eps (z , \bxi_2 ,\bzeta_2; \tau) 
= \hat{\eta} \Big(\frac{z}{\eps} ,\bxi_2,\bzeta_2 ; \tau\Big) 
\exp \Big( i \frac{z}{\eps k_o } \bxi_2\cdot \bzeta_2\Big).
\end{equation}
In the regime (\ref{eq:r1}) the rescaled function $\tilde{\eta}^\eps$ 
satisfies the equation with fast phases
\begin{equation}
\label{eq:tildemueps}
 \frac{\partial \tilde{\eta}^\eps}{\partial z}  ={\cal L}^\eps_z \tilde{\eta}^\eps, 
 \end{equation}
 where 
\ba
&& \nonumber{\cal L}^\eps_z  \tilde{\eta}  (  \bxi_2,  \bzeta_2)=
\frac{\sigma_{\rm m}^2 k_o^2 \ell_{\rm m}^3}{4 (2\pi)^2} 
\int_{\RR^2} \hat{C}(\bk \ell_{\rm m}) \bigg[  - 2\tilde{\eta} (   \bxi_2 ,\bzeta_2) +
2 \tilde{\eta}  (  \bxi_2-\bk,  \bzeta_2) 
e^{i\frac{z}{\eps k_o} \bk \cdot \bzeta_2 }
  \\
\nonumber
&& \hbox{}
+ 
2 g_\tau \tilde{\eta}  ( \bxi_2,  \bzeta_2-\bk) 
e^{i\frac{z}{\eps k_o} \bk \cdot  \bxi_2  } 
- g_\tau
\tilde{\eta} (   \bxi_2-\bk , \bzeta_2-\bk) 
 e^{i\frac{z}{\eps k_o} (  \bk \cdot (\bzeta_2+\bxi_2)-|\bk|^2 )} \\
 && \hbox{}
- g_\tau \tilde{\eta}  ( \bxi_2-\bk,  \bzeta_2+\bk) 
e^{i\frac{z}{\eps k_o} ( \bk \cdot (\bzeta_2-\bxi_2)+|\bk|^2)}
\bigg] d \bk , \hspace*{0.2in}
\label{eq:tildeNeps}
\ea
and the initial condition is
\begin{equation}
\label{eq:initialtildeM2eps}
\tilde{\eta}^\eps(z=0,\bxi_2,  \bzeta_2; \tau ) = (2\pi)^4 
\phi_{\eps/\ell_{\rm s}} ( \bxi_2 ) \delta ( \bzeta_2 ) +
(2\pi)^4 f_\tau^2
 \phi_{\eps/\ell_{\rm s}} ( \bzeta_2 ) \delta ( \bxi_2 ) .
\end{equation}
 Note that $\phi_{\kappa}$ belongs to $L^1$ and has a $L^1$-norm equal to one.
The asymptotic behavior as $\eps \to 0$ of the moments is therefore determined
by the solutions of partial differential equations with rapid phase terms.
We can now proceed as in \cite{gar16a} and we obtain the following proposition.
\begin{prop}
\label{prop:sci1}%
In the regime (\ref{eq:r1}), 
the function $\tilde{\eta}^\eps(z, \bxi_2 ,\bzeta_2 ; \tau) $ has the form
\ba
\label{eq:propsci11}
&&  \tilde{\eta}^\eps(z, \bxi_2 ,\bzeta_2 ; \tau)  = 
   K(z)
\phi_{\eps/\ell_{\rm s}} ( \bxi_2 )
\delta ( \bzeta_2 ) 
+
K(z) A_1 (z,\bxi_2,{\bf 0}) \delta(\bzeta_2)
\\
\nonumber
&& 
 \hbox{} +
K(z) A_{g_\tau} \big(z,\bzeta_2, \frac{\bxi_2}{\eps}\big) \phi_{\eps/\ell_{\rm s}}(\bxi_2)
+  f_\tau^2 K(z)
\delta( \bxi_2 )
\phi_{\eps/\ell_{\rm s}} ( \bzeta_2 ) \\
& &  \hbox{} +
 f_\tau^2 K(z)A_{g_\tau} (z,\bzeta_2,{\bf 0}) \delta(\bxi_2) \nonumber
+
 f_\tau^2 K(z) A_{1} \big(z,\bxi_2, \frac{\bzeta_2}{\eps}\big) \phi_{\eps/\ell_{\rm s}}(\bzeta_2)
 \\
& &  \hbox{} 
 + R^\eps  (z , \bxi_2 , \bzeta_2; \tau )   , \nonumber
\ea
where the functions $K$ and $A_g$  are defined by
\begin{align}
\label{def:K}
K(z)  =& (2\pi)^4 \exp\Big(- \frac{\sigma_{\rm m}^2 k_o^2 \ell_{\rm m} z}{2} 
 \Big) , \\
\nonumber
A_g (z,\bxi,\bzeta)   =&  \frac{1}{(2\pi)^2}
 \int_{\RR^2}  \Big[  \exp \Big( \frac{\sigma_{\rm m}^2 k_o^2 \ell_{\rm m} g}{2} \int_0^z C\Big(  \frac{\bx}{\ell_{\rm m}} + \frac{\bzeta z' }{k_o \ell_{\rm m} } \big) dz' \Big) -1\Big]\\
 & \times
   \exp \big( -i \bxi\cdot \bx  \big)
 d\bx  ,   
\label{def:A}
\end{align}
and the function $R^\eps $ satisfies
$
\sup_{z \in [0,L]} \| R^\eps (z,\cdot,\cdot ; \tau) \|_{L^1(\RR^2\times \RR^2 )} 
\stackrel{\eps \to 0}{\longrightarrow}  0 $.
\end{prop}
We remark that 
\be
 \frac{ K^{1/4}(z)}{2\pi}  =  \exp\Big(- \frac{\sigma_{\rm m}^2 k_o^2 \ell_{\rm m} z }{8} 
 \Big) 
\ee
represents damping of the mean   wave field  due to scattering and 
transfer of coherent energy to incoherent wave energy in the case with frozen medium 
and deterministic sources.   
We remark moreover that the factor  $A_g$ depends on the two point statistics of the random medium at lateral offsets and
captures effects of lateral  scattering of wave field energy.    
As a result, the quantity of interest in (\ref{eq:expressmu4eta}) is then 
\begin{equation}
\mu_4^\eps(L,{\bf 0},{\bf 0},{\bf 0},{\bf 0}; \tau) = 1 +f_\tau^2  \exp \Big(- \frac{\sigma_{\rm m}^2  (1-g_\tau)  k_o^2 \ell_{\rm m} 
 L}{2}\Big)
+
{\cal Q}_{g_\tau}(L) 
+
 f_\tau^2  {\cal Q}_{1}(L) 
\end{equation}
with
\begin{align}
\nonumber
{\cal Q}_g(L) =& 
\exp \Big(- \frac{\sigma_{\rm m}^2 k_o^2 \ell_{\rm m}
L}{2}\Big)  \int_{\RR^2} \phi_{\ell_{\rm s}^{-1}}(\bzeta) \\
&\times \Big[ 
 \exp \Big( \frac{\sigma_{\rm m}^2 k_o^2 \ell_{\rm m} g }{2} \int_0^L  C\big( \frac{\bzeta z}{k_o \ell_{\rm m}}  \big)  d z \Big)-1\Big]  d\bzeta  .
\end{align}
Therefore the relative covariance of the intensities at time zero  and time $\tau$ is
\ban
  \frac{\mu_4^\eps(L,{\bf 0},{\bf 0},{\bf 0},{\bf 0}; \tau) - \mu_2^\eps(L,{\bf 0},{\bf 0} ; 0)^2}
{ \mu_2^\eps(L,{\bf 0},{\bf 0} ; 0)^2}
&=& f_\tau^2  \exp \Big(- \frac{ \sigma_{\rm m}^2 (1-g_\tau) k_o^2  \ell_{\rm m}
L}{2}\Big)
\\ && 
%\hspace*{4cm} 
 \hbox{}  +
{\cal Q}_{g_\tau}(L)  
+
 f_\tau^2  {\cal Q}_{1}(L) 
 .
\ean
This gives the result (\ref{eq:S1}) for the scintillation index in the regime $\ell_{\rm s} \gg \ell_{\rm m}$.    
  
\subsection{Scintillation Regime  with an Intermediate Correlation Radius of the Source}
\label{sec:proof2}%
We consider the white-noise paraxial regime in which, additionally, the correlation radius of the source is of the same order as 
the correlation radius of the medium $\ell_{\rm s}\sim \ell_{\rm m}$.  This is the regime when the source lateral spatial fluctuations takes place on the same scale of variation as that of the
random microstructure fluctuations, rather 
than being large relative to this scale as in the previous Section \ref{sec:proof1}.
More exactly, we here deal with the following scaled regime:
\begin{equation}
\label{eq:r2}
    \frac{\ell_{\rm m}}{\ell_{\rm s}} \sim 1 \, ,  \quad\quad
     \frac{L}{\ell_{\rm s}} \sim   \alpha^{-1} \eps^{-1} \, ,   \quad\quad
     \frac{\lambda_o}{\ell_{\rm s}} \sim \alpha   \,   ,    \quad\quad
     \sigma_{\rm m}^2 \sim \alpha^3 \eps \,      ,
\end{equation}
and we assume $  \alpha \ll \eps \ll 1$ [Note that $L/ \ell_{\rm mfp} \sim 1$ and $\lambda_o L /\ell_{\rm m}^2 \sim \eps^{-1}$].
This means that the paraxial white-noise limit $\alpha \to 0$ is taken first, and then we want to apply the limit  $\eps\to 0$ in the fourth-moment equation (\ref{eq:fouriermom0eta}).
% in order to  derive the result (\ref{eq:S2}).  
%
%We introduce a small dimensionless parameter $\eps$ and we now rescale
%as:
%\begin{equation}
%\ell_{\rm s} \to  {\ell_{\rm s}} ,\quad C(\bx) \to \eps C(\bx), \quad L \to \frac{L}{\eps}.
%\label{eq:r2}
%\end{equation}
As above we  introduce  the rescaled function
\begin{equation}
\tilde{\eta}^\eps (z , \bxi_2 ,\bzeta_2; \tau) = \hat{\eta} \Big(\frac{z}{\eps} ,\bxi_2,\bzeta_2 ; \tau\Big) 
\exp \Big( i \frac{z}{k_o\eps} \bxi_2\cdot \bzeta_2\Big).
\end{equation}
In the  regime (\ref{eq:r2}) the rescaled function $\tilde{\eta}^\eps$  
satisfies  again  the equation with fast phases (\ref{eq:tildemueps}-\ref{eq:tildeNeps}), 
here with  initial condition given by (\ref{eq:4ic}).  
 The asymptotic behavior as $\eps \to 0$ of the moments is therefore determined
by the solutions of partial differential equations with rapid phase terms.
We can again proceed similarly as  in \cite{gar16a} and we obtain the following proposition.
 \begin{prop}
\label{prop:sci12}%
In the scintillation regime (\ref{eq:r2}), 
the function $\tilde{\eta}^\eps(z, \bxi_2 ,\bzeta_2 ; \tau) $ has the form
\begin{equation}
 \tilde{\eta}^\eps(z, \bxi_2 ,\bzeta_2 ; \tau)  =
(2\pi)^4  
B_1 (z,\bxi_2 ) \delta(\bzeta_2)
+
(2\pi)^4 f_\tau^2 B_{g_\tau} (z, \bzeta_2 ) \delta(\bxi_2)  
+ R^\eps  (z , \bxi_2 , \bzeta_2; \tau )   ,
\label{eq:propsci12}
\end{equation}
with 
\begin{equation}
B_g (z,\bxi ) =
\frac{1}{(2\pi)^2} \int_{\RR^2} \exp\Big( - i\bxi \cdot\bx - \frac{|\bx|^2}{2\ell_{\rm s}^2} - \frac{\sigma_{\rm m}^2 k_o^2 \ell_{\rm m} z}{2} \big[
%C({\bf 0})
1 - g C( \frac{\bx }{ \ell_{\rm m}}) \big]\Big) d\bx,
\end{equation}
and the function $R^\eps $ satisfies
$
\sup_{z \in [0,L]} \| R^\eps (z,\cdot,\cdot ; \tau) \|_{L^1(\RR^2\times \RR^2 )} 
\stackrel{\eps \to 0}{\longrightarrow}  0 $.
\end{prop}

As a result, the quantity of interest (\ref{eq:expressmu4eta}) is
\begin{equation}
\mu_4^\eps(L,{\bf 0},{\bf 0},{\bf 0},{\bf 0}; \tau) =  1+
 f_\tau^2 
 \exp \Big(- \frac{\sigma_{\rm m}^2 (1-g_\tau) k_o^2  \ell_{\rm m}
 L }{2}    \Big)
\end{equation}
and
\begin{equation}
\frac{\mu_4^\eps(L,{\bf 0},{\bf 0},{\bf 0},{\bf 0}; \tau) - \mu_2^\eps(L,{\bf 0},{\bf 0} ; 0 )^2}
{ \mu_2^\eps(L,{\bf 0},{\bf 0} ; 0 )^2}
=
 f_\tau^2 
 \exp \Big(- \frac{\sigma_{\rm m}^2 (1-g_\tau) k_o^2 \ell_{\rm m}
 L }{2}    \Big)
 .
 \end{equation}
This then gives the result (\ref{eq:S2}) for the scintillation index.

\subsection{Scintillation Regime with a Small Correlation Radius of the Source}
\label{sec:proof3}%
We finally consider the white-noise paraxial regime in which, additionally, the correlation radius of the source is smaller than 
the correlation radius of the medium $\ell_{\rm s}\ll \ell_{\rm m}$ and derive the result (\ref{eq:S3}) for the scintillation index in this regime. 
More exactly, we here deal with the following scaled regime:
\begin{equation}
    \frac{\ell_{\rm m}}{\ell_{\rm s}} \sim \eps^{-1} \, ,  \quad\quad
     \frac{L}{\ell_{\rm s}} \sim   \alpha^{-1}
     % \eps^{-1} 
      \, ,   \quad\quad
     \frac{\lambda_o}{\ell_{\rm s}} \sim \alpha %\eps^{-1}
       \,   ,    \quad\quad
     \sigma_{\rm m}^2 \sim \alpha^3 \eps^{-1} \,      ,
     \label{eq:r3}
\end{equation}
and we assume $  \alpha \ll \eps \ll 1$
 [Note that $L/ \ell_{\rm mfp} \sim \eps^{-2}$ and $\lambda_o L /\ell_{\rm s}^2 \sim 1$].
This means that the paraxial white-noise limit $\alpha \to 0$ is taken first, and then we want to apply the limit  $\eps\to 0$ in the fourth-moment equation (\ref{eq:fouriermom0eta}).
%  
%We introduce a small dimensionless parameter $\eps$ and  rescale in this case  as:
% \begin{equation}
%\ell_{\rm s} \to \ell_{\rm s}, \quad  C(\bx) \to \eps^{-2} C(\eps \bx) ,\quad L \to L.
%\label{eq:r3}
%\end{equation}
We also assume that $C$ is smooth and isotropic, so that we have (\ref{eq:ex}),
and also 
$\frac{1}{(2\pi)^2} \int_{\RR^2} \hat{C}(\bq) \bq \otimes \bq d\bq = 2c_2 {\bf I} $.
We denote by $\hat{\eta}^\eps$ the function (\ref{eq:defhateta}) in the regime (\ref{eq:r3}).
Then we find that 
in the regime of small $\eps$ 
the function $\hat{\eta}^\eps(z,\bxi_2,\bzeta_2; \tau)$ is solution to the system 
\begin{equation}\label{eq:d}
\frac{\partial \hat{\eta}^\eps}{\partial z} + \frac{i}{k_o}    \bxi_2\cdot \bzeta_2  \hat{\eta}^\eps
=
\frac{\sigma_{\rm m}^2 k_o^2  c_2 (1-g_\tau)}{2 \ell_{\rm m}} \Delta_{\bxi_2} \hat{\eta}^\eps ,
\end{equation}
 with   initial condition given by (\ref{eq:4ic}). 
 We can easily solve (\ref{eq:d})  via a Fourier transform and find  
\begin{prop}
\label{prop:sci2}%
In the scintillation regime (\ref{eq:r3}), 
the function $\hat{\eta}^\eps(z, \bxi_2 ,\bzeta_2 ; \tau) $ 
has the form
\begin{equation}
\hat{\eta}^\eps(z, \bxi_2 ,\bzeta_2 ; \tau)  =
(2\pi)^4  
G_1 (z,\bxi_2; \tau) \delta(\bzeta_2)
+
(2\pi)^4 f_\tau^2 G_2 (z,\bxi_2,\bzeta_2; \tau) \phi_{1/\ell_{\rm s}} (\bzeta_2)  ,
% + R^\eps  (z , \bxi_2 , \bzeta_2; \tau )   ,
\end{equation}
with
\begin{align}
G_1 (z,\bxi_2; \tau) =&
\frac{\ell_{\rm s}^2}{2\pi} \frac{1}{1+ \frac{\sigma_{\rm m}^2 (1-g_\tau) c_2 k_o^2 \ell_{\rm s}^2 L}{\ell_{\rm m}}}
\exp\Big( - \frac{\ell_{\rm s}^2 |\bxi_2|^2}{2\big( 1+ \frac{\sigma_{\rm m}^2 (1-g_\tau) c_2 k_o^2 \ell_{\rm s}^2 L}{\ell_{\rm m}} \big)}\Big) ,\\
\nonumber
G_2 (z,\bxi_2,\bzeta_2; \tau) =& 
\frac{1}{(2\pi)^2}
\int_{\RR^2}
\exp \Big( -\frac{c_2 (1-g_\tau) \sigma_{\rm m}^2 k_o^2 }{2\ell_{\rm m}} \int_0^L \big| \bx - \frac{\bzeta_2 z}{k_o}\big|^2 dz-i \bxi_2 \cdot \bx \Big)
 d\bx \\
 \nonumber
 =&
 \frac{\ell_{\rm m}}{2\pi  c_2  (1-g_\tau) \sigma_{\rm m}^2 k_o^2 L} \exp \Big(
 - 
 \frac{iL}{2k_o} \bxi_2\cdot \bzeta_2  \Big)
 \\
 & \times
 \exp\Big( -\frac{c_2 (1-g_\tau) \sigma_{\rm m}^2 L^3}{24 \ell_{\rm m}} |\bzeta_2|^2 
 -
 \frac{\ell_{\rm m}}{2c_2 (1-g_\tau) \sigma_{\rm m}^2 k_o^2  L} |\bxi_2|^2 \Big) 
   .
\end{align}
%and $ \sup_{z \in [0,L]} \| R^\eps (z,\cdot,\cdot ) \|_{L^1(\RR^2\times \RR^2 )} 
%\stackrel{\eps \to 0}{\longrightarrow}  0  $.
\end{prop}

As a result
\begin{equation}
\mu_4^\eps(L,{\bf 0},{\bf 0},{\bf 0},{\bf 0}; \tau) = 1  +
\frac{ f_\tau^2 }{1+ 
\frac{  c_2 (1-g_\tau) \sigma_{\rm m}^2 L^3}{3 \ell_{\rm m}  \ell_{\rm s}^2}}
\end{equation}
and
\begin{equation}
\frac{\mu_4^\eps(L,{\bf 0},{\bf 0},{\bf 0},{\bf 0}; \tau) - \mu_2^\eps(L,{\bf 0},{\bf 0} ; 0)^2}
{ \mu_2^\eps(L,{\bf 0},{\bf 0} ; 0)^2}
= 
\frac{ f_\tau^2 }{1+ 
\frac{ c_2 (1-g_\tau)  \sigma_{\rm m}^2 L^3}{3 \ell_{\rm m} \ell_{\rm s}^2}}  .
\end{equation}
This then gives  (\ref{eq:S3}) for the scintillation index when $\ell_{\rm s} \ll \ell_{\rm m}$.
  
%\begin{backmatter}
\section*{Funding}
 JG was supported by the Agence Nationale pour la Recherche under Grant No. ANR-19-CE46-0007 (project ICCI), 
 and Air Force Office of Scientific Research under grant FA9550-18-1-0217.
 \\
KS was supported by the Air Force Office of Scientific Research under grant FA9550-18-1-0217,  and   the National Science Foundation under grant DMS-2010046.

\section*{Acknowledgments}
 We thank C.~Nelson, S.~Avramov-Zamurovic, O.~Korotkova, S.~Guth and R.~Malek-Madani 
  for allowing us to use their data from \cite{svetlana}  to generate
 Figure \ref{fig:9}. 
  
%\bmsection{Disclosures}
% The authors declare no conflicts of interest.
% 
%\bmsection{Data availability} Data underlying the results presented in Figure \ref{fig:9} was published  in
%\cite{svetlana}.
%  No other data was analyzed in the presented research and no data was generated.
  
%\bmsection{Supplemental document}
%See Supplement 1 for supporting content. 

%\end{backmatter}

%%%%%%%%%% If preparing manually:

\end{document}